\documentclass[aps,pra,superscriptaddress,twocolumn,longbibliography]{revtex4-2}

\usepackage[utf8]{inputenc}
\usepackage{mathtools,amsmath}
\usepackage{amsfonts,mathtools}
\usepackage{amssymb}
\usepackage{comment}
\usepackage{color}
\usepackage[colorlinks=true,
			urlcolor=blue,
			citecolor=blue,
			linkcolor=blue,
			citebordercolor={1 0 0},
			linkbordercolor={0 0 1}]{hyperref}
\usepackage{amsthm}
\usepackage{graphicx}
\usepackage[normalem]{ulem} %%for strikethrough - use command \sout{ }
\usepackage{dsfont}

\usepackage{ragged2e}
\usepackage{caption}
\usepackage{subcaption}
\usepackage{array}
\usepackage{enumerate}
\usepackage{cleveref}

\usepackage{float}
\makeatletter
\let\newfloat\newfloat@ltx
\makeatother
\usepackage{algorithm,algorithmic}
\newcommand{\figref}[1]{Fig.~\ref{#1}}

\newcommand{\tr}{{\rm Tr}}

\newcommand{\bea}{\begin{eqnarray}}
\newcommand{\eea}{\end{eqnarray}}
\newcommand{\be}{\begin{equation}}
\newcommand{\ee}{\end{equation}}
\newcommand{\ba}{\begin{equation}\begin{aligned}}
\newcommand{\ea}{\end{aligned}\end{equation}}
\newcommand{\eqs}[1]{\begin{align}#1\end{align}}

\newcommand{\red}{\textcolor{red}}
\newcommand{\blue}{\textcolor{blue}}

\newtheorem{theorem}{Theorem}

\newtheorem*{definition*}{Definition}
\newtheorem*{remark*}{Remark}

\theoremstyle{definition}

\def\1{\mathds{1}}

\def\mD{\mathcal{D}}
\def\mE{\mathcal{E}}

\def\mM{\mathcal{M}}
\def\mN{\mathcal{N}}

\def\mR{\mathcal{R}}

\def\mU{\mathcal{U}}

\def\ml{\mathfrak{L}}

\def\({\left(}
\def\){\right)}
\def\[{\left[}
\def\]{\right]}

\makeatletter
\let\Hy@backout\@gobble
\makeatother
\begin{document}

\title{Practical limitations of quantum data propagation on noisy quantum processors}
	
 %%authors
\author{Gaurav Saxena}
\email{gaurav.saxena@lge.com}	
\affiliation{LG Electronics Toronto AI Lab, Toronto, Ontario M5V 1M3, Canada} 

\author{Ahmed Shalabi}
\affiliation{LG Electronics Toronto AI Lab, Toronto, Ontario M5V 1M3, Canada}

\author{Thi Ha Kyaw}
\affiliation{LG Electronics Toronto AI Lab, Toronto, Ontario M5V 1M3, Canada}

%\author{other authors?}
%\email{other.author@email.com}
%\affiliation{some other university}
\date{\today}
	
\begin{abstract}
The variational quantum imaginary time evolution algorithm is efficient in finding the ground state of a quantum Hamiltonian.
This algorithm involves solving a system of linear equations in a classical computer and the solution is then used to propagate a quantum wavefunction. 
Here, we show that owing to the noisy nature of current quantum processors, such a quantum algorithm or the family of quantum algorithms will require single- and two-qubit gates with very low error probability to produce reliable results.
Failure to meet such condition will result in erroneous quantum data propagation even for a relatively small quantum circuit ansatz.
Specifically, we provide the upper bounds on how the relative error in variational parameters' propagation scales with the probability of noise in quantum hardware.
%, for partially dephasing noise\sout{errors} and for partially depolarizing noise \sout{errors}} 
We also present an exact expression of how the relative error in variational parameter propagation scales with the probability of partially depolarizing noise.
%\sout{Our work challenges the mainstream notion of hybrid quantum-classical quantum algorithms being able to perform under noisy environments while we show such algorithms in fact require very low error quantum gates to get reliable results.}
%\red{\sout{Our work indicates that extra attention is required in propagating variational parameters by using measurement results from noisy quantum hardware.}}
\end{abstract}

\maketitle
%{
%  \hypersetup{linkcolor=black}
%  \tableofcontents
%}
\newpage

%{\textit{Introduction.---}}
\section{Introduction}
Quantum computers promise a quantum advantage in numerous applications such as factoring \cite{Shor1994Nov}, solving linear systems~\cite{Harrow2009Oct}, quantum simulations \cite{Georgescu2014Mar}, and quantum chemistry/material discovery \cite{Cao2019Oct}. 
The aforementioned quantum applications sometimes require millions of physical qubits and gates \cite{Lee2021Jul}, which current quantum hardware does not possess \cite{Lin2022Feb} and hence hinders the direct experience of any proposed practical quantum speedup \cite{TorstenHoefler2023May}.
In the current noisy intermediate-scale quantum (NISQ) era \cite{Preskill2018Jan}, many interesting practical applications are evaluated through hybrid quantum-classical algorithms otherwise known as NISQ algorithms \cite{Bharti_2022,Tilly2022Nov,Cerezo2021Sep}.
These algorithms are designed for short-depth quantum circuits with limited numbers of qubits and gates.
One such important class of algorithms are variational quantum algorithms (VQAs) \cite{Peruzzo2014Jul,McClean2016Feb,Wecker2015Oct}.
VQAs use short-depth parameterized quantum circuits to perform some quantum computation and then leverage classical computing power for optimization in a feedback fashion. 
It is, however, an open question whether NISQ computers will yield a quantum advantage for practically meaningful problems. 

\begin{figure*}[th]
    \centering
    \includegraphics[scale=1.0]{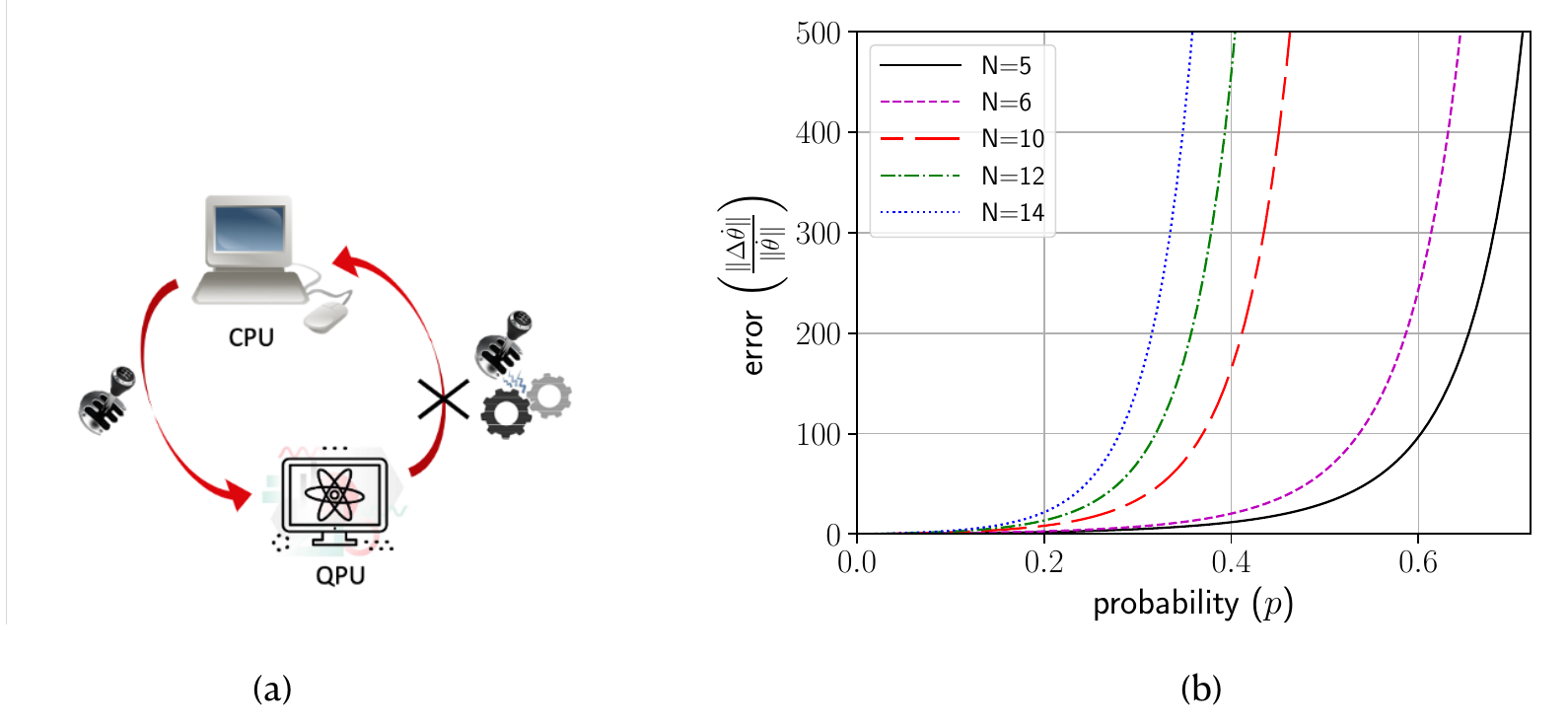}
    \caption{\justifying{(a) A simple illustration to show that owing to noisy nature of the current quantum processors, the outcomes from the QPU are misestimated and thus cannot be relied upon {without any error mitigation}.
    (b) Plot of the upper bound on the relative error in the rate of change of the parameters with respect to the probability with which the depolarizing noise occurs for different ansatzes with same condition number but different number of parameters $N$, when $N=6$ and $N=14$ can be seen as simulating H$_2$ and LiH molecules in the minimum basis set using two layers of the hardware-efficient ansatz \cite{Kandala2017Sep}, respectively. }}
    \label{fig:defining-image}
\end{figure*}

One significant family of VQA algorithms is the quantum imaginary time evolution (QITE)~\cite{Motta_2019,McArdle_2019, Nishi_2021}.
QITE finds its uses in various applications including efficiently finding the ground state of a Hamiltonian, diffusion quantum Monte Carlo, etc~\cite{McArdle_2019,Motta_2019,Nishi_2021, RevModPhys.73.33, Yeter-Aydeniz2020Jul, Tsuchimochi2022May, Yuan2019Oct,PhysRevLett.125.010501, Gomes2020Oct}.
QITE works by propagating quantum data to evolve a quantum state.
In general, imaginary time evolution or a Wick rotation~\cite{PhysRev.96.1124} is a simple but powerful mathematical technique that involves mapping to the time coordinate $\tau = i t $.
While simulating imaginary time evolution can be done classically, the resources required scale exponentially with the size of the Hilbert space in question \cite{Lee2022Aug}.
Performing \emph{quantum} imaginary time evolution involves non-unitary evolution and thus, implementing it using digital quantum computers is hard as quantum processors only allow for unitary operations via quantum gates.
To address this limitation, a hybrid quantum-classical variational approach was proposed to implement the quantum imaginary time evolution~\cite{McArdle_2019,Motta_2019,Nishi_2021}.
Imaginary time evolution of the Wick-rotated Schr{\"o}dinger equation is implemented by encoding an initial quantum state as a parameterized quantum state $|\varphi(\tau)\rangle \approx|\phi(\theta(\tau))\rangle$ where $\theta$ is a real valued parameter vector $\theta(\tau)=\left(\theta_1(\tau), \theta_2(\tau), \ldots, \theta_{N}(\tau)\right)$.
To simulate non-unitary dynamics, {an approximate} formulation of imaginary time evolution derived from a generalized McLachlan’s variational principle is used to evolve the parameterized variational circuit~\cite{Yuan2019Oct}.
The final step is to solve a linear equation of the form, (see \figref{fig:defining-image}(a) and {section} \ref{sec:qite_mixedstates})
\begin{equation}
    M \dot{\mathbf{\theta}} = Y\, .
\nonumber 
\end{equation}
%\sout{which involves inverting the matrix $M$ to solve for $\dot{\theta}$, which}
{The solution $\dot{\theta}$} then dictates the evolution of variational parameters of the ansatz circuit for the next time step: $\theta(\tau+\Delta \tau)\approx \theta(\tau)+ \dot{\theta}\times \Delta \tau$.
The matrix $M$ and the vector $Y$ depend on the ansatz circuit and the given Hamiltonian.
Similarly, for the real-time evolution, we have the same equation where $Y$ is replaced by a different vector $V$ and the equation becomes $M \dot{\mathbf{\theta}} = V$ (see Section \ref{sec:qite_mixedstates} for details).

Even though the family of QITE algorithms is well-studied in the literature and various improved algorithms of the same kind have been developed/proposed recently~\cite{Kyaw2022Aug,Zhang2023Jan, PhysRevResearch.3.033083}, none of the works have assessed~\cite{zoufal2021error,StilckFranca2021Nov, McArdle_2019, Wang2021Nov} its performance on noisy quantum processors.
Given the current noisy nature of quantum processors, it becomes crucial to understand how well the quantum imaginary (real) time evolution algorithm performs in absence of any error correction or mitigation technique.
In particular, we are interested in the impact of hardware errors incurred during each iteration on the exact rate of change of parameters, $\dot{\theta}$ used to estimate the next set of variational parameters $\theta(\tau+\Delta \tau)$.

In this {Article}, we provide the quantitative answer to the above question by providing upper bounds on the error incurred in exactly solving $M\dot{\theta}=Y$ (see \figref{fig:defining-image}(b)).
{This paper is arranged as follows. First, in Section~\ref{sec:qite_mixedstates}, we provide a brief review to the quantum imaginary time evolution for completeness.
Then in Section~\ref{sec:error_models}, we give the details of our error model. Section~\ref{sec:main_results} presents our main results followed by Discussion in Section \ref{sec:discuss}.}

\section{The Quantum Imaginary Time Evolution Algorithm for mixed states}\label{sec:qite_mixedstates}
In this section, we will present an overview of the QITE algorithm for mixed states. 
For a more comprehensive overview, we refer the reader to Ref.~\cite{Yuan2019Oct} and citations therein. Before we begin, a quick aside on notation. $R_k$ denotes the $k$-th rotation gate (which is a single or two-qubit gate) and $\mathcal{R}_k$ denotes the $k$-th quantum gate acting on the whole system.  $\mR_k(\tau)$ on an operator $\tau$ is to be understood as the action of $R_k$ gate on $\tau$ by conjugation, i.e., $R_k\tau R_k^{\dagger}$. $R_k\tau$ is to be understood as the multiplication of the operators.

Under imaginary time evolution where the imaginary time $\tau := it$, an initial mixed state $\rho(0)$ evolves according to
\eqs{
\rho(\tau)=\frac{e^{-H \tau} \rho(0) e^{-H \tau}}{\operatorname{Tr}\left[e^{-2 H \tau} \rho(0)\right]}.
}
The above time evolution can be rewritten as 
\eqs{\frac{d \rho(\tau)}{d \tau}=-(\{H, \rho(\tau)\}-2\langle H\rangle \rho(\tau)).}
Encoding the state $\rho$ in a parameterized circuit gives us the following time evolution equation for a parameterized mixed state $\rho(\theta(\tau))$
\eqs{
\sum_i \frac{\partial \rho(\vec{\theta})}{\partial \theta_k} \dot{\theta}_k=-(\{H, \rho(\vec{\theta}(\tau))\}-2\langle H\rangle \rho(\vec{\theta}(\tau))).
}

The evolution of the $\theta$ parameters can be derived by calculating the variation $\delta \rho (\theta)$ and using the MacLachlan's variational principle 
\eqs{
\delta\|d \rho / d \tau+(\{H, \rho\}-2\langle H\rangle \rho)\|=0,
}
which yields the following matrix equation for the time evolution of $\vec{\theta}$
\eqs{
\sum_q M_{k, q} \dot{\theta}_q=Y_k.
}
The matrix elements of $M$ are defined as
\eqs{
M_{k, q}:=\operatorname{Tr}\left[\left(\frac{\partial \rho(\vec{\theta}(t))}{\partial \theta_k}\right)^{\dagger} \frac{\partial \rho(\vec{\theta}(t))}{\partial \theta_q}\right],
}
and for $Y$
\eqs{
Y_k:=-\operatorname{Tr}\left[\left(\frac{\partial \rho(\vec{\theta})}{\partial \theta_k}\right)^{\dagger}(\{H, \rho(\vec{\theta}(t))\}-2\langle H\rangle \rho(\vec{\theta}(t)))\right]
}
It is important to note that when $\vec{\theta}$ is real, $M$ and $Y$ are also real and $Y$ will simplify to
\eqs{
Y_k=-\operatorname{Tr}\left[\frac{\partial \rho(\vec{\theta}(t))}{\partial \theta_k}\{H, \rho(\vec{\theta}(t))\}\right]
}
To implement the imaginary time evolution for mixed states, we need to measure the coefficients $M_{k,q}$ and $Y_k$. We utilize the general circuit shown in Fig.~\ref{fig:M_kq_circuit} to measure $\frac{\partial \rho}{\partial \theta_k}$. The generator of the circuit can be expressed as
\eqs{
\mathcal{L}(\rho)=\sum_k g_k S_k \rho T_k^{\dagger}
}
where $S_k$ and $T_k$ are unitary operators and $g_k$ are coefficients. Using the above we can write
\eqs{
\frac{\partial \mathcal{R}_k(\rho)}{\partial \lambda_k}=\sum_i r_{k, i} S_{k, i} \rho T_{k, i}^{\dagger}.
}
From there we can rewrite the derivatives $\frac{\partial \rho}{\partial \theta_k}$ to obtain the following expressions for $M$ and $Y$~\cite{Yuan2019Oct}\footnote{Note two minor corrections in Fig.~3 of Ref.~\cite{Yuan2019Oct}. First one is that $U_k:= S_k^{\dagger}T_k$, and the second is that in the second system, $T_q$ will come after $U_q$ and not $S_q$ (as given in Fig.~3 of \cite{Yuan2019Oct}) as the definition of $M_{k,q}$ involves taking the conjugate of one term.}:
\eqs{
\label{eqs:M_kq_defn}
\begin{split}
    M_{k,q} &= \sum_{i,j} \frac{1}{2}\{ r^*_{k,i}r_{q,j}\tr\left[\left(\mR_{k,i}\rho_0\right)^{\dagger}\left(\mR_{q,j}\rho_0 \right)\right] + \rm{h.c.}, \\
Y_k &= - \sum_{i}\frac{1}{2}\{r_{k,i} \tr\[(\mR_{k,i}\rho_0)(\rho H) \] \}
\end{split}
 }
where $\rho$ is the final state given by 
\eqs{
\rho = \mR_N\circ \mR_{N-1} \circ \cdots \circ \mR_1(\rho_0)
}
and
\eqs{
\mR_{k,i}\rho_0 = \mR_N\cdots \mR_{k+1}[S_{k,i}\left(\mR_{k-1}\cdots \mR_2 \mR_1(\rho_0)\right) T_{k,i}^{\dagger}]\,.}

Further, to simulate the real time evolution the same setup can be used~\cite{Yuan2019Oct}.
The main difference in the real and imaginary time evolution is in the matrix equation to be solved where $Y$ is replaced by $V$ for the real evolution, i.e., the linear equation for matrix inversion becomes $M\dot{\theta} = V$ where V is given by 
\eqs{
\label{V_k}
V_k=\sum_{i, j} \frac{1}{2}\left\{r_{k, i} g_j \operatorname{Tr}\left[\left(\mathcal{R}_{k, i} \rho_0\right)\left(S_j \rho T_j^{\dagger}\right)\right]+\text { h.c. }\right\}.
}

%\newpage
%{\textit{Error models.---}}
\section{Error Models}\label{sec:error_models}
\begin{figure*}[t]
    \centering
    \includegraphics[scale=0.75]{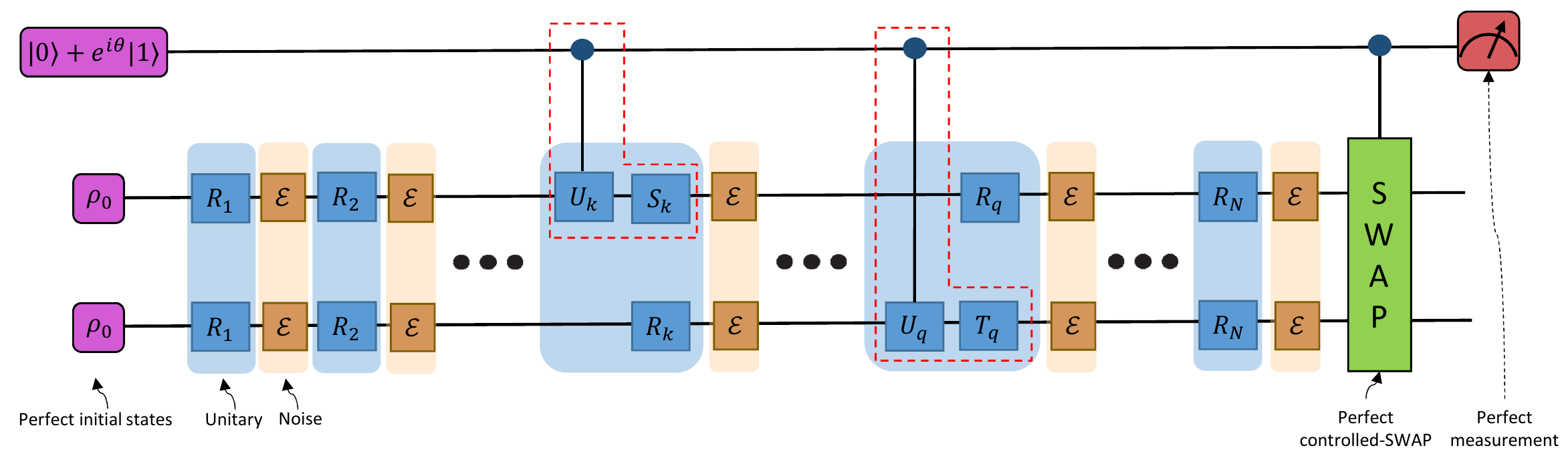}
    \caption{\justifying{Error setting in our studies.
    The above circuit evaluates the elements of the matrix $M$ (see Sec.~\ref{sec:qite_mixedstates} for the form of $M_{k,q}$ element) with global error $\mE$ occurring after every gate where $\mE(\rho):= p\mN(\rho) + (1-p)\rho$. 
    The above circuit can be decomposed into three parts. The first part/line is a single qubit (in the state $|0\rangle + e^{i\theta}|1\rangle$) that controls the circuits below.
    The second and third parts/lines each have the same set of gates and the same number of multiple qubits, and represent the same ansatz circuits.
    The difference in the second and third lines at some sites (here, $k$-th and $q$-th site) arise because of the differentiation with respect to the respective parameter at those sites which is required for computing the matrix element. For representing the differentiation with respect to the respective parameter, the gate at that site is replaced by an appropriate controlled-unitary followed by appropriate unitary at that site.
    In addition, we consider ansatz circuits with only single and two-qubit gates.
    We are assuming that the initial state preparation, the first control qubit, and the measurements are error-free (fuzzy-free in the diagram). 
    For each application of controlled-$U_k$($U_q$) followed by $S_k$($T_q$) as the $k$($q$)-th gate in the circuit, the error is applied after the combined gates (red dashed box). 
     }}
    \label{fig:M_kq_circuit}
\end{figure*}

We consider the errors of the following probabilistic form:
\eqs{\label{eq:noise_model}
\mE(\rho):= p\mN(\rho) + (1-p)\rho}
where $\mN$ is some quantum channel, representing the hardware error occurring with probability $p$ in the quantum circuit.
We assume that the error $\mE$ occurs after every gate implementation and is a global {Markovian} error (see \figref{fig:M_kq_circuit}).
Considering such general Markovian error allows us to estimate the error scaling in the worst-case scenario. 
Such error models are commonly studied in literature, however most studies include local errors on all qubits and not global ~\cite{Xue2021Mar, Marshall2020Aug, Wang2021Nov}. By considering global errors, we also account for cases where noise model could be correlated.
Note that in this article, we do not consider temporal correlations, and thus have avoided all non-Markovian errors in this analysis.
We further impose a physical condition that the channel $\mN\in {\rm CPTP}(A_0\to A_1)$ is such that $\|\mN(X)\|_2\leq \|X\|_2 \;\forall\, X\in \ml(A_0)$, where $\|\cdot\|_2$ denotes the Frobenius norm~\footnote{We denote $\|\cdot\|_2$ and $\|\cdot\|$ interchangeably to indicate the Frobenius norm defined as $\|A\|_2 = \sqrt{{\rm Tr}(A^* A)}$},
%, whereas $\|\cdot\|$ is used to indicate the matrix 2-norm defined as $\|A\| = \sqrt{\lambda_{\rm max}A^* A}$.}, 
$\ml(A_0)$ denotes the set of all linear operators in the Hilbert space $A_0$, and CPTP$(A_0 \to A_1) $\footnote{We denote a system and its corresponding Hilbert space using an uppercase letter with a numerical subscript, like $A_0$ or $A_1$.} denotes the set of completely positive and trace preserving maps that map operators in the Hilbert space $A_0$ to operators in the Hilbert space $A_1$.
%\sout{We required this extra condition to prove our bounds mathematically to corroborate practical limitations of variational quantum imaginary (real) time evolution.}
{With this condition, we also leave the algorithmic analyses due to the presence of replacement channels along with non-Markovian noise} \cite{Bastidas2018Sep,Kyaw2017May,Gour2019Mar,Regula2021Jul} as an open problem.
Notice that the set of all channels $\mN$ that obey $\|\mN(X)\|_2\leq \|X\|_2 \;\forall\, X\in \ml(A_0)$ contain the set of all probabilistic unitary channels, i.e., channels of the form
\eqs{\mN(\rho) = \sum_{i} p_i \mathcal{U}_i(\rho) = \sum_{i} p_i U_i \rho U_i^{\dagger}\,}
where $p_i \geq 0$ and $\sum_i p_i =1$.
By doing so, we are accounting for all the probabilistic unitary errors that contain the set of random Pauli errors. 
The special cases of which are the (partially and completely) dephasing and depolarizing channels, which are considered very relevant experimentally \cite{Mi2022Nov, PhysRevA.104.062432}. %\sout{and the main foci of our paper's results}.

%{\textit{Error bounds.---}}
\section{Error Bounds}\label{sec:main_results}

By assuming the initial state preparation and measurements to be error-free (see \figref{fig:M_kq_circuit}) and by considering gate errors as described above in Section~\ref{sec:error_models}, {one can obtain upper bounds on the error $\epsilon$ (defined as the norm of the difference between the ideal and erroneous $\dot{\theta}$ in the exact solution of $M\dot{\theta} = Y$) using fundamental inequalities of perturbation theory (see Eq.~\eqref{eq:initial_UB} of the appendix). 
However, the bound obtained this way is very loose and blows out very quickly, thereby giving any sensible bound only for very small probabilities of noise.
}

\begin{figure*}[t]
    \centering
    \begin{subfigure}[t]{0.5\textwidth}
        \centering
        \includegraphics[scale=0.55]{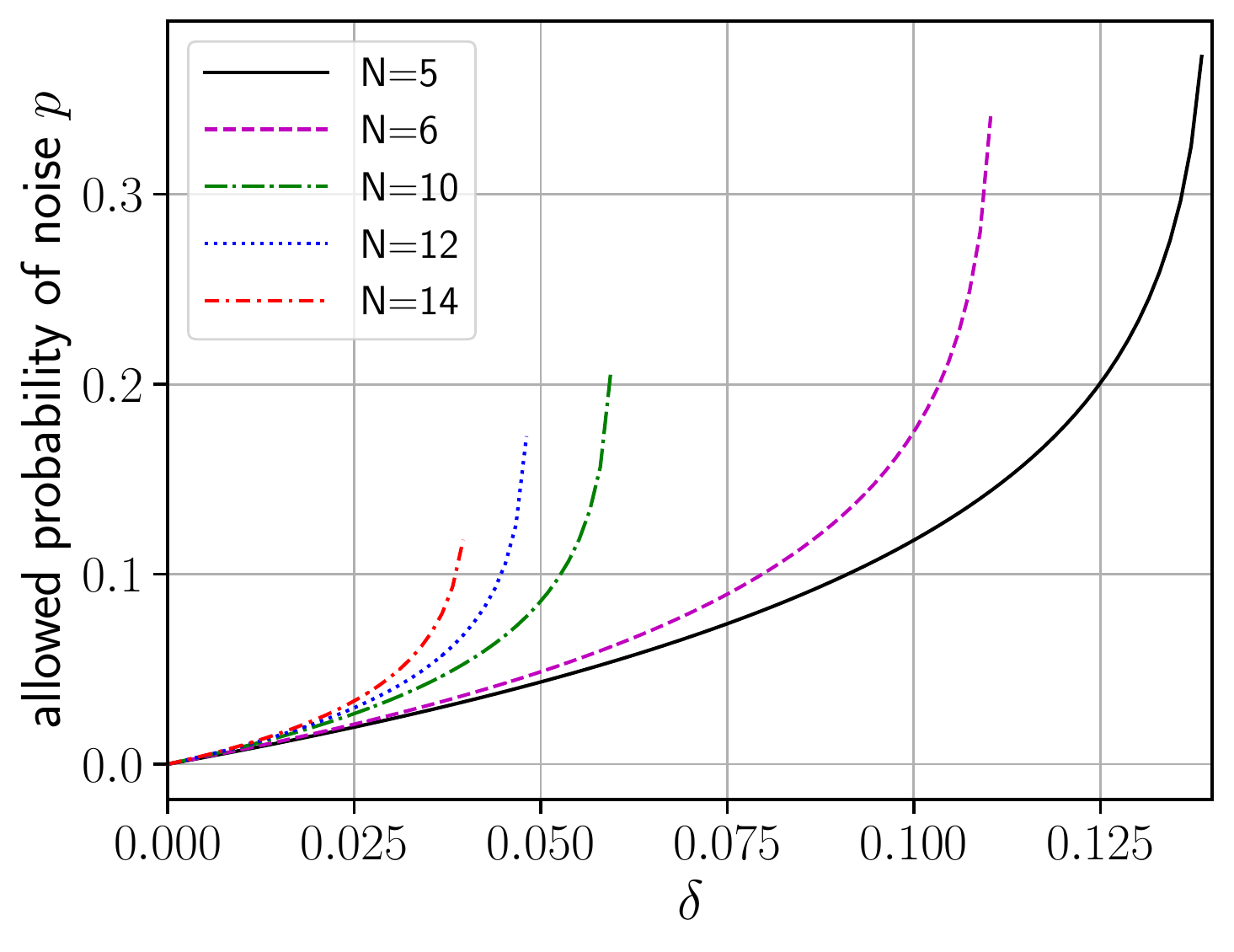}
        \caption{}
        \label{fig:delta_vs_prob}
    \end{subfigure}%
    ~ 
    \begin{subfigure}[t]{0.5\textwidth}
        \centering
        \includegraphics[scale =0.55]{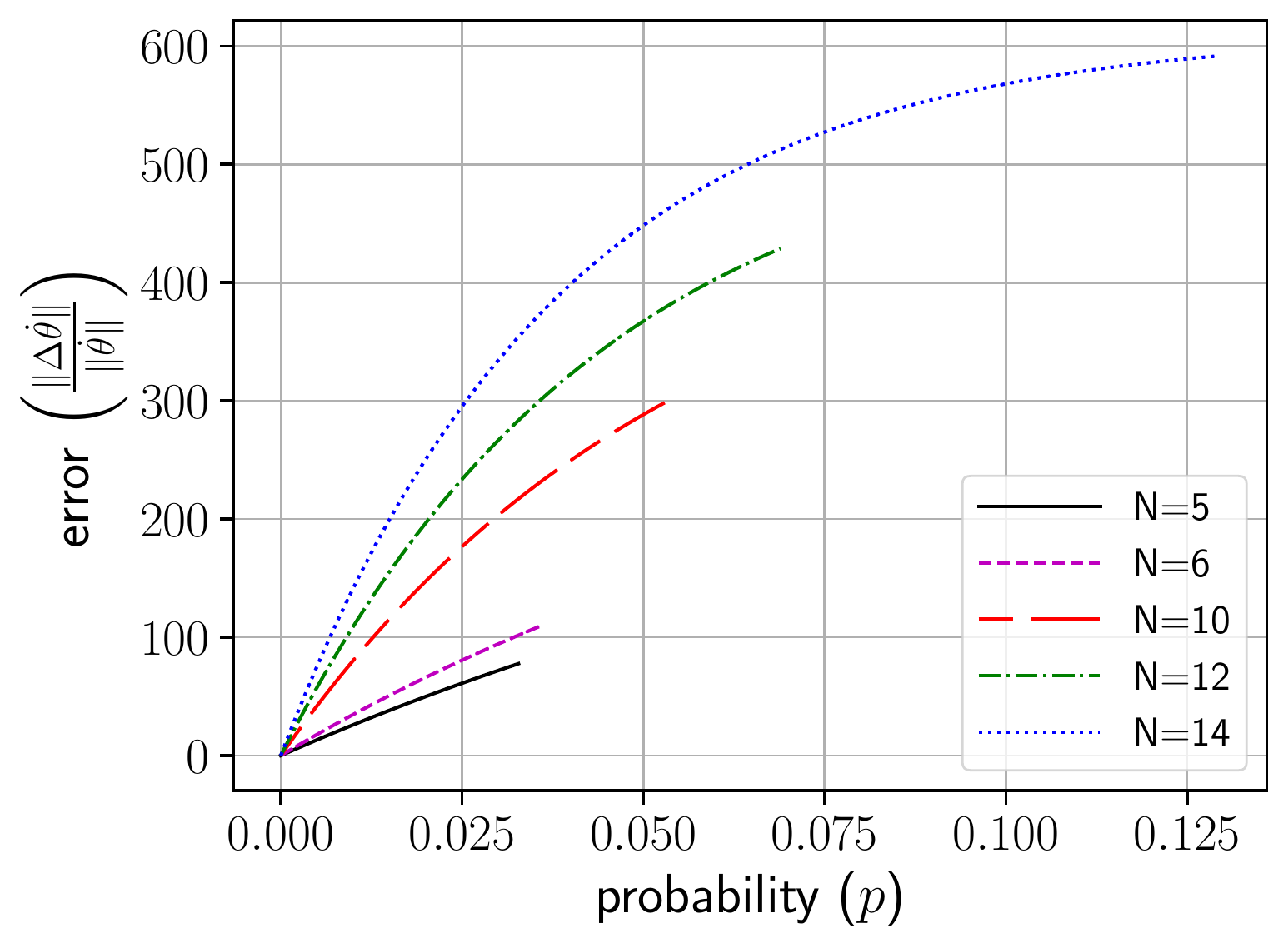}
        \caption{}
        \label{fig:err_vs_prob_5params_tightbound}
    \end{subfigure}
    \caption{\justifying{{Plots for Theorem 1 results. In the above plots, we fixed the condition number of $M$ and $\|M\|$ to be $66.7$ and $0.9977$, respectively, and varied the number of parameters. These values were obtained by using the five-parameter ansatz considered for simulating the Hamiltonian of the Hydrogen molecule and by generating random angles (see Appendix~\ref{app:numerics} for details). 
    Figure (a) is the plot of the constraint in Eq.~\eqref{eq:constraint_on_prob_tight_ub}. In this graph, we plotted the allowed maximum probability with respect to fixed $\delta$ which dictates when $\Delta M \Delta \dot{\theta}$ can be considered negligible. For each given $N$, the graph stops after a certain $\delta$; it shows that beyond that $\delta$ the right hand side in Eq.~\eqref{eq:constraint_on_prob_tight_ub} will be negative, thus leading to invalid bound. To plot Fig. (b), we choose $\delta = 0.04$ for which we calculated the allowed probabilities to be $0.032, 0.036, 0.052, 0.068, 0.129$ for $N = 5, 6, 10, 12, 14$, respectively. Figure (b) is the plot of the relative error in $\dot{\theta}$ as a function of probability of noise. The plot stops at the maximum allowed probability as governed from the constraint. 
    }}}
    \label{fig:plots_thm1_fiveparam}
\end{figure*}
{Tight upper bounds can be found by imposing stricter conditions on solving the linear equations or by knowing more details of the noise.
We consider instances of both these cases and find tight upper bounds.}

{When we solve the perturbed linear equation $M\dot{\theta} = Y$, we are essentially solving the equation $(M + \Delta M)(\dot{\theta} + \Delta \dot{\theta}) = Y + \Delta Y$.
Now for a fixed very small $\delta <<1$, if each element of $\Delta M$ and $\Delta \dot{\theta}$ is less than $\delta$, i.e., $(\Delta M)_{i,j}\leq \delta$ and $(\dot{\theta})_j\leq \delta$, then $(\Delta M \Delta \dot{\theta})_i \sim O(\delta^2)$ and can safely be ignored as its contribution to $\Delta Y$ can be considered negligible.
By slightly loosening this restriction, we can ask for $\| \Delta M\|\leq N^2 \delta$ and $\|\Delta \dot{\theta}\|\leq N \delta$ to hold for the contribution of $\Delta M \Delta \dot{\theta}$ to be considered negligible. 
Considering these constraints, we get the following tight upper bound for the general noise model.}
{
\begin{theorem}\label{thm:tight_upper_bound_gen_error}
For probabilistic errors $\mE$ of the form:
$\mE(\rho) = p \mN(\rho) + (1-p) \rho$, where the channel $\mN$ is such that $\|\mN(X)\|_2\leq \|X\|_2 \;\forall\, X\in \ml(\mathcal{H})$ where $\ml$ denotes the set of linear operators in a Hilbert space $\mathcal{H}$, we show that the upper bound in the error $\epsilon := \|\Delta\dot{\theta}\|$ is given by
\begin{widetext}
    \eqs{
\frac{\epsilon}{\|\dot{\theta} \|} \leq {\rm cond}(M)
\left(\left(1- (1-p)^{2N}\right)\left( 1+\frac{N}{2\|M\|} \right) + (1- (1-p)^N)\left( 1 + \frac{\sqrt{N}}{\sqrt{2\|Y\|}}\right)\right)
}
\end{widetext}
for the probability of noise $p$ in the following range
\eqs{\label{eq:constraint_on_prob_tight_ub}0\leq p\leq 1- \left( 1- \frac{2N^2\delta}{N+2\|M\|}\right)^{\frac{1}{2N}}
}
where $\delta$ in the term on the right is small enough such that the expression is positive.
\end{theorem}
}
{The proof of the above theorem has been provided in Appendix~\ref{app:tighter_UB_proof}.
Note that beyond the probability range provided in Eq.~\eqref{eq:constraint_on_prob_tight_ub}, the contribution of $\Delta M \Delta \dot{\theta}$ to $\Delta Y$ cannot be considered negligible.
We also numerically demonstrate the constraint of Theorem 1 in Fig.~\ref{fig:plots_thm1_fiveparam}(a), where we show how the allowed probability (for which the main result of Theorem 1 is valid) changes with varying $\delta$ and number of parameters, $N$. 
By fixing $\delta = 0.04$, we show in Fig.~\ref{fig:plots_thm1_fiveparam}(b) how the relative error scales with the probability of noise in quantum hardware. Note that for smaller $N$, we can have larger $\delta$ which will lead to larger allowed probabilities.
}

Now let us consider the evolution of quantum state in QITE under a specific noise model, the partially depolarizing noise, which is a major source of error in NISQ computers~\cite{PhysRevA.104.062432}.
For the depolarizing noise, we use its symmetry to find the relation between the relative error, the probability of noise, and the number of parameters.

\begin{theorem}\label{thm:upper_bound_depolarizing}
For partially depolarizing channels, the upper bound in the error $\epsilon := \|\Delta\dot{\theta}\|$ is given by
\eqs{
\frac{\epsilon}{\|\dot{\theta}\|}  = \frac{1-(1-p)^N}{(1-p)^N}.\label{eq:th2}
}

\end{theorem}

\noindent The proof of this theorem can be found in Appendix \ref{app:C}.
There are several key features to note here.
First, {we now have an exact expression on how the error in the solution of $M\dot{\theta} = Y$ will scale as opposed to Theorem 1 where we only had an upper bound.
The above exact expression is because (in Appendix~\ref{app:C}) we were able to derive exactly the form of $M^{err}$ as a function of $M$ (see Eq.~\eqref{Merr_thm2}). 
%\sout{The upper bound in Eq.~\eqref{eq:th2} comes purely from trying to invert a perturbed matrix \eqref{eqn:CondNumbIneq}.} 
 Moreover, the relative error under depolarizing noise only depends on the probability of depolarizing noise and the number of parameters $N$ and is independent of the vector $Y$ or the matrix $M$, i.e. it is independent of the Hamiltonian being simulated or the structure of the ansatz circuit.}
 
{It is clear from the expression that as the probability of noise increases, the error also increases.
This leads us to conclude that QITE can perform well only when the probability of noise is very low, or in other words, under depolarizing noise, QITE in fact requires very high fidelity quantum gates without any error mitigation deployed.
Moreover, analyzing the behaviour of error scaling under the depolarizing noise, we can conjecture that as the probability increases (see Fig. 1(b)), the error scaling for general noise model must also scale up rapidly with increasing probability of noise. 
Lastly, given the exact expression of relative error, we can perform a simple post-processing (mitigation) to get the accurate $\dot{\theta}$.}

{From the two theorems above, we see that as the noise probability after each gate increases, the relative error $\epsilon/\| \dot{\theta}\|$ grows very rapidly or diverges. This means that as the noise probability increases, the new parameters to be updated that are required for the next run of the QITE will be further away from the actual parameters and so the error in estimating the ground state of a given Hamiltonian will increase.}
Also note that if we consider circuits where multiple unitary gates are applied onto different qubit registers simultaneously, and the probabilistic error is applied after each such layer, then the {expressions of relative errors} are similar to the {results in Theorem 1 and 2} except that the number of parameters, $N$, in the equations will be replaced by the depth, $d$, of the circuit.
%\red{\sout{Moreover, the upper bounds on the error in the evolution of parameters for real time evolution will be similar to the above bounds (of the imaginary time evolution) except some prefactors,  and $Y$ will be replaced with $V$}} \sout{(see appendix for the expression of $V$)}.

In addition, we perform numerical analyses of our results by considering the paradigmatic Hydrogen molecule H$_2$ in the minimal STO-3G basis. 
The reduced H$_2$ qubit Hamiltonian in the Bravyi-Kitaev representation is given by \cite{McArdle_2019}
\eqs{H = g_0I + g_1 Z_0 + g_2 Z_1 + g_3Z_0Z_1 + g_4Y_0 Y_1 + g_5 X_0X_1\,}
where $g_0 = 0.2252$, $g_1 = 0.3435$,
$g_2 = -0.4347$,
$g_3 = 0.5716$, $g_4 = 0.0910$, and 
$g_5 = 0.0910$.
In our analysis, we make use of two ansatzes to understand our error bounds.
One is a simple ansatz and the other ansatz is the universal ansatz~\cite{Peruzzo2014Jul} as shown in Figs.~\ref{fig:ansatz}(a) and~\ref{fig:ansatz}(b) of the Appendix, respectively.
Notice that the simple ansatz has 5 parameters and the universal ansatz has 9 parameters.
Using these ansatzes, we then plot the error vs probability figures throughout. 
For example, \figref{fig:defining-image}(b) was obtained by using the 5-parameter simple ansatz circuit undergoing depolarizing noise to get the reference line $(N=5,$ black solid line$)$ with random angles to generate one matrix $M$ with a reference condition number, cond$(M)$. 
While fixing the same cond$(M)$ {and $\|M\|$}, we can now start to vary the number of parameters $N$.
By doing so, we effectively change the structure of the ansatz circuit.
Alternatively, when $N<5$, we can set some $\theta=0$ and vary the other parameters to get the same condition number while using the same ansatz circuit.
The same strategy is applied for the general probabilistic errors (Fig.~\ref{fig:plots_thm1_fiveparam}(b) of the main text and Fig.~\ref{fig:plots_err_vs_p} of Appendix \ref{app:numerics}), and for the depolarizing error (Fig.~\ref{fig:defining-image}(b) of the main text or Fig.~\ref{fig:err_vs_p_thm2} of Appendix \ref{app:numerics}).
From all of the presented numerical results, {we find that the error in updating the variational parameters increases drastically with an increase in error probability in the physical hardware. Since depolarizing noise commonly occurs in superconducting qubits architecture, we expect that such behaviour could become a bottleneck in performing QITE in high-depth quantum circuits without appropriate post processing or error mitigation.} %\sout{one can find that the error in finding the rate of change of parameters increases drastically even with a minute increase in error probability in the physical quantum hardware, which is the direct consequence of our theorems.}

Finally, it is important to note that one does not have to solve, $M\dot{\theta} = Y$ via matrix inversion or any method in particular. 
One can use any numerical method to solve the linear system of equations. Our goal is to understand how the relative error in $\dot{\theta}$ scales with the probability of noise in quantum hardware.
In doing so, we found the upper bound of relative error and showed that it is directly proportional to the condition number of $M$ (see Eq.(17)). 
The condition number tells us how sensitive the solution of a linear solver is given a slight perturbation in the matrix $M$ or vector $Y$.
Now, concerning the computation of condition number of $M$, computing the norm of the inverse/pseudo inverse of $M$ might be difficult, specially for large matrices. 
However, it is possible to approximate the condition number of a matrix without computing the norm of its inverse~\cite{tobias_lect}, thus resolving the issue of computing the inverse/condition number of a matrix.

%{\textit{Discussions.---}}
\section{Discussion}\label{sec:discuss}

In this work we presented upper bounds on the error incurred during a single iteration of quantum imaginary time evolution in the presence of global Markovian noise in the quantum circuit.
We assumed that the error can occur after every gate in the ansatz circuit with a probability $p$.
Furthermore, we derived an exact expression for the error incurred specifically due to partially depolarizing channels.
We note that these error analyses are also applicable to real-time quantum data evolution. This is because the variational problem for real time evolution is similar since the matrix M in the linear problem of solving for $\dot{\theta}$ is exactly the same. i.e., the results from the two theorems would not differ except to a different dependence on the norm of $V$ (equation \eqref{V_k}).
Our results primarily show that to implement the variational quantum data evolution both real- and imaginary-time evolution, we would require gates with very low error probability due to the sensitivity of matrix inversion to external environmental perturbations.
We also show that scalability is a problem with this approach because of how the error scales with the number of parameters in the circuit and the probability of gate errors.
In fact, when we consider multiple unitary gates applying onto different qubit registries, $N$, the notation we used for the number of variational parameters, can be taken as the quantum circuit depth. 
Hence, the results in \figref{fig:defining-image} still hold with $N$ then being the depth of the circuit.
Furthermore, our results can be used to compare different ansatzes in order to choose a well-suited ansatz for the given problem.
In other words, if the amount of error that can be tolerated is fixed, one can find the maximum probability of allowed noise in the system and thus, relevant error mitigation techniques can be employed in order to restrict error probability going beyond a certain threshold.
%\sout{Similarly, if the probability of noise occurrence is known, then the ansatz for which the error is less can be considered for the QITE algorithm.}

This work is important considering the rising interest in extracting quantum advantage from NISQ algorithms and processors and which makes it critical to do an in-depth error analysis in order to truly understand whether the NISQ algorithms really stand a chance in showing quantum advantage.
We found that the error analyses performed on QITE algorithm, which as we stated previously is one of the most efficient NISQ algorithms to find the ground state of a Hamiltonian, were either too basic, or assumed very strict conditions, or were based on simple numerical simulations~\cite{Kyaw2022Aug,Zhang2023Jan, PhysRevResearch.3.033083, zoufal2021error,McArdle_2019, Wang2021Nov}. 
For instance, in Ref.~\cite{zoufal2021error}, the error analysis presented was done by considering two different evolutions of the state and how those two evolutions diverged. No hardware noise was considered in getting the theoretical results. In Ref.~\cite{McArdle_2019}, no specific hardware noise or even general error model was considered. By assuming $p=10^{-4}$ error rate per gate, the authors numerically concluded that ``the imaginary time algorithm can perform significantly better than the gradient descent". However, our analysis is much more general.
To show this let us take the same numbers as considered in Ref.~\cite{McArdle_2019}, and consider that the probability of depolarizing noise after every gate is $10^{-4}$.
Our results (shown in Fig.\ref{fig:defining-image}(b)) suggest that the error incurred would be very small and so, with this noise probability, we can infer that the imaginary time evolution algorithm will perform well - which is in agreement with the results of Ref.~\cite{McArdle_2019}.

Finally, it is yet to be seen how various quantum error mitigation techniques \cite{Cai2022Oct,Quek2022Oct} would help to suppress error in {variational parameters propagation from solving a linear system of equations as we have presented here}.  
We conjecture that the standard zero noise extrapolation method \cite{Temme2017Nov,Li2017Jun,Kandala2019Mar,Giurgica-Tiron2020Oct} would fail due to the need of multiple quantum circuit layers for its extrapolation technique to work.
However, we expect probabilistic error cancellation (PEC) \cite{Temme2017Nov,Endo2018Jul,Song2019Sep,Zhang2020Jan} might help to reduce the growth of errors in the evolution of the circuit parameters.
%\sout{The tug-of-war between SEC and PEC is left to be seen in future work.}
{Exact details on the amount of error suppression with respect to the overhead needed for PEC are left for a future work}.
%The impact and potential improvement that PEC might yield is left to be seen in future work. 

%%%%%%%%%%%%%%%%%%%%%%%%%%%%%%%%%%%%%%%%%%%%%%%%%%%%%%%%%%%%%%%
%\textit{Code availability.---} 
\section*{Code Availability}
Code and data generated to construct the plots will be made available upon request.
%%%%%%%%%%%%%%%%%%%%%%%%%%%%%%%%%%%%%%%%%%%%%%%%%%%%%%%%%%%%%%%
\section*{Acknowledgements}
We thank Kevin Ferreira (Senior Director of LG Electronics Toronto AI Lab) for his constant support throughout this work. %We also would like to thank the anonymous QIP2024 referees for their reviews.

\bibliography{ref}

\begin{thebibliography}{55}%
\makeatletter
\providecommand \@ifxundefined [1]{%
 \@ifx{#1\undefined}
}%
\providecommand \@ifnum [1]{%
 \ifnum #1\expandafter \@firstoftwo
 \else \expandafter \@secondoftwo
 \fi
}%
\providecommand \@ifx [1]{%
 \ifx #1\expandafter \@firstoftwo
 \else \expandafter \@secondoftwo
 \fi
}%
\providecommand \natexlab [1]{#1}%
\providecommand \enquote  [1]{``#1''}%
\providecommand \bibnamefont  [1]{#1}%
\providecommand \bibfnamefont [1]{#1}%
\providecommand \citenamefont [1]{#1}%
\providecommand \href@noop [0]{\@secondoftwo}%
\providecommand \href [0]{\begingroup \@sanitize@url \@href}%
\providecommand \@href[1]{\@@startlink{#1}\@@href}%
\providecommand \@@href[1]{\endgroup#1\@@endlink}%
\providecommand \@sanitize@url [0]{\catcode `\\12\catcode `\$12\catcode `\&12\catcode `\#12\catcode `\^12\catcode `\_12\catcode `\%12\relax}%
\providecommand \@@startlink[1]{}%
\providecommand \@@endlink[0]{}%
\providecommand \url  [0]{\begingroup\@sanitize@url \@url }%
\providecommand \@url [1]{\endgroup\@href {#1}{\urlprefix }}%
\providecommand \urlprefix  [0]{URL }%
\providecommand \Eprint [0]{\href }%
\providecommand \doibase [0]{https://doi.org/}%
\providecommand \selectlanguage [0]{\@gobble}%
\providecommand \bibinfo  [0]{\@secondoftwo}%
\providecommand \bibfield  [0]{\@secondoftwo}%
\providecommand \translation [1]{[#1]}%
\providecommand \BibitemOpen [0]{}%
\providecommand \bibitemStop [0]{}%
\providecommand \bibitemNoStop [0]{.\EOS\space}%
\providecommand \EOS [0]{\spacefactor3000\relax}%
\providecommand \BibitemShut  [1]{\csname bibitem#1\endcsname}%
\let\auto@bib@innerbib\@empty
%</preamble>
\bibitem [{\citenamefont {Shor}(1994)}]{Shor1994Nov}%
  \BibitemOpen
  \bibfield  {author} {\bibinfo {author} {\bibfnamefont {P.}~\bibnamefont {Shor}},\ }\bibfield  {title} {\bibinfo {title} {Algorithms for quantum computation: discrete logarithms and factoring},\ }in\ \href {https://doi.org/10.1109/SFCS.1994.365700} {\emph {\bibinfo {booktitle} {Proceedings 35th Annual Symposium on Foundations of Computer Science}}}\ (\bibinfo {address} {Santa Fe, NM, USA},\ \bibinfo {year} {1994})\ pp.\ \bibinfo {pages} {124--134}\BibitemShut {NoStop}%
\bibitem [{\citenamefont {Harrow}\ \emph {et~al.}(2009)\citenamefont {Harrow}, \citenamefont {Hassidim},\ and\ \citenamefont {Lloyd}}]{Harrow2009Oct}%
  \BibitemOpen
  \bibfield  {author} {\bibinfo {author} {\bibfnamefont {A.~W.}\ \bibnamefont {Harrow}}, \bibinfo {author} {\bibfnamefont {A.}~\bibnamefont {Hassidim}},\ and\ \bibinfo {author} {\bibfnamefont {S.}~\bibnamefont {Lloyd}},\ }\bibfield  {title} {\bibinfo {title} {{Quantum Algorithm for Linear Systems of Equations}},\ }\href {https://doi.org/10.1103/PhysRevLett.103.150502} {\bibfield  {journal} {\bibinfo  {journal} {Phys. Rev. Lett.}\ }\textbf {\bibinfo {volume} {103}},\ \bibinfo {pages} {150502} (\bibinfo {year} {2009})}\BibitemShut {NoStop}%
\bibitem [{\citenamefont {Georgescu}\ \emph {et~al.}(2014)\citenamefont {Georgescu}, \citenamefont {Ashhab},\ and\ \citenamefont {Nori}}]{Georgescu2014Mar}%
  \BibitemOpen
  \bibfield  {author} {\bibinfo {author} {\bibfnamefont {I.~M.}\ \bibnamefont {Georgescu}}, \bibinfo {author} {\bibfnamefont {S.}~\bibnamefont {Ashhab}},\ and\ \bibinfo {author} {\bibfnamefont {F.}~\bibnamefont {Nori}},\ }\bibfield  {title} {\bibinfo {title} {{Quantum simulation}},\ }\href {https://doi.org/10.1103/RevModPhys.86.153} {\bibfield  {journal} {\bibinfo  {journal} {Rev. Mod. Phys.}\ }\textbf {\bibinfo {volume} {86}},\ \bibinfo {pages} {153} (\bibinfo {year} {2014})}\BibitemShut {NoStop}%
\bibitem [{\citenamefont {Cao}\ \emph {et~al.}(2019)\citenamefont {Cao}, \citenamefont {Romero}, \citenamefont {Olson}, \citenamefont {Degroote}, \citenamefont {Johnson}, \citenamefont {Kieferov{\ifmmode\acute{a}\else\'{a}\fi}}, \citenamefont {Kivlichan}, \citenamefont {Menke}, \citenamefont {Peropadre}, \citenamefont {Sawaya} \emph {et~al.}}]{Cao2019Oct}%
  \BibitemOpen
  \bibfield  {author} {\bibinfo {author} {\bibfnamefont {Y.}~\bibnamefont {Cao}}, \bibinfo {author} {\bibfnamefont {J.}~\bibnamefont {Romero}}, \bibinfo {author} {\bibfnamefont {J.~P.}\ \bibnamefont {Olson}}, \bibinfo {author} {\bibfnamefont {M.}~\bibnamefont {Degroote}}, \bibinfo {author} {\bibfnamefont {P.~D.}\ \bibnamefont {Johnson}}, \bibinfo {author} {\bibfnamefont {M.}~\bibnamefont {Kieferov{\ifmmode\acute{a}\else\'{a}\fi}}}, \bibinfo {author} {\bibfnamefont {I.~D.}\ \bibnamefont {Kivlichan}}, \bibinfo {author} {\bibfnamefont {T.}~\bibnamefont {Menke}}, \bibinfo {author} {\bibfnamefont {B.}~\bibnamefont {Peropadre}}, \bibinfo {author} {\bibfnamefont {N.~P.~D.}\ \bibnamefont {Sawaya}}, \emph {et~al.},\ }\bibfield  {title} {\bibinfo {title} {{Quantum Chemistry in the Age of Quantum Computing}},\ }\href {https://doi.org/10.1021/acs.chemrev.8b00803} {\bibfield  {journal} {\bibinfo  {journal} {Chem. Rev.}\ }\textbf {\bibinfo {volume} {119}},\ \bibinfo {pages} {10856} (\bibinfo {year} {2019})}\BibitemShut
  {NoStop}%
\bibitem [{\citenamefont {Lee}\ \emph {et~al.}(2021)\citenamefont {Lee}, \citenamefont {Berry}, \citenamefont {Gidney}, \citenamefont {Huggins}, \citenamefont {McClean}, \citenamefont {Wiebe},\ and\ \citenamefont {Babbush}}]{Lee2021Jul}%
  \BibitemOpen
  \bibfield  {author} {\bibinfo {author} {\bibfnamefont {J.}~\bibnamefont {Lee}}, \bibinfo {author} {\bibfnamefont {D.~W.}\ \bibnamefont {Berry}}, \bibinfo {author} {\bibfnamefont {C.}~\bibnamefont {Gidney}}, \bibinfo {author} {\bibfnamefont {W.~J.}\ \bibnamefont {Huggins}}, \bibinfo {author} {\bibfnamefont {J.~R.}\ \bibnamefont {McClean}}, \bibinfo {author} {\bibfnamefont {N.}~\bibnamefont {Wiebe}},\ and\ \bibinfo {author} {\bibfnamefont {R.}~\bibnamefont {Babbush}},\ }\bibfield  {title} {\bibinfo {title} {{Even More Efficient Quantum Computations of Chemistry Through Tensor Hypercontraction}},\ }\href {https://doi.org/10.1103/PRXQuantum.2.030305} {\bibfield  {journal} {\bibinfo  {journal} {PRX Quantum}\ }\textbf {\bibinfo {volume} {2}},\ \bibinfo {pages} {030305} (\bibinfo {year} {2021})}\BibitemShut {NoStop}%
\bibitem [{\citenamefont {Lin}\ and\ \citenamefont {Tong}(2022)}]{Lin2022Feb}%
  \BibitemOpen
  \bibfield  {author} {\bibinfo {author} {\bibfnamefont {L.}~\bibnamefont {Lin}}\ and\ \bibinfo {author} {\bibfnamefont {Y.}~\bibnamefont {Tong}},\ }\bibfield  {title} {\bibinfo {title} {{Heisenberg-Limited Ground-State Energy Estimation for Early Fault-Tolerant Quantum Computers}},\ }\href {https://doi.org/10.1103/PRXQuantum.3.010318} {\bibfield  {journal} {\bibinfo  {journal} {PRX Quantum}\ }\textbf {\bibinfo {volume} {3}},\ \bibinfo {pages} {010318} (\bibinfo {year} {2022})}\BibitemShut {NoStop}%
\bibitem [{\citenamefont {Hoefler}\ \emph {et~al.}(2023)\citenamefont {Hoefler}, \citenamefont {H\"{a}ner},\ and\ \citenamefont {Troyer}}]{TorstenHoefler2023May}%
  \BibitemOpen
  \bibfield  {author} {\bibinfo {author} {\bibfnamefont {T.}~\bibnamefont {Hoefler}}, \bibinfo {author} {\bibfnamefont {T.}~\bibnamefont {H\"{a}ner}},\ and\ \bibinfo {author} {\bibfnamefont {M.}~\bibnamefont {Troyer}},\ }\bibfield  {title} {\bibinfo {title} {Disentangling hype from practicality: On realistically achieving quantum advantage},\ }\href {https://doi.org/10.1145/3571725} {\bibfield  {journal} {\bibinfo  {journal} {Commun. ACM}\ }\textbf {\bibinfo {volume} {66}},\ \bibinfo {pages} {82–87} (\bibinfo {year} {2023})}\BibitemShut {NoStop}%
\bibitem [{\citenamefont {Preskill}(2018)}]{Preskill2018Jan}%
  \BibitemOpen
  \bibfield  {author} {\bibinfo {author} {\bibfnamefont {J.}~\bibnamefont {Preskill}},\ }\bibfield  {title} {\bibinfo {title} {{Quantum Computing in the NISQ era and beyond}},\ }\href {https://doi.org/10.22331/q-2018-08-06-79} {\bibfield  {journal} {\bibinfo  {journal} {Quantum}\ }\textbf {\bibinfo {volume} {2}},\ \bibinfo {pages} {79} (\bibinfo {year} {2018})}\BibitemShut {NoStop}%
\bibitem [{\citenamefont {Bharti}\ \emph {et~al.}(2022)\citenamefont {Bharti}, \citenamefont {Cervera-Lierta}, \citenamefont {Kyaw}, \citenamefont {Haug}, \citenamefont {Alperin-Lea}, \citenamefont {Anand}, \citenamefont {Degroote}, \citenamefont {Heimonen}, \citenamefont {Kottmann}, \citenamefont {Menke} \emph {et~al.}}]{Bharti_2022}%
  \BibitemOpen
  \bibfield  {author} {\bibinfo {author} {\bibfnamefont {K.}~\bibnamefont {Bharti}}, \bibinfo {author} {\bibfnamefont {A.}~\bibnamefont {Cervera-Lierta}}, \bibinfo {author} {\bibfnamefont {T.~H.}\ \bibnamefont {Kyaw}}, \bibinfo {author} {\bibfnamefont {T.}~\bibnamefont {Haug}}, \bibinfo {author} {\bibfnamefont {S.}~\bibnamefont {Alperin-Lea}}, \bibinfo {author} {\bibfnamefont {A.}~\bibnamefont {Anand}}, \bibinfo {author} {\bibfnamefont {M.}~\bibnamefont {Degroote}}, \bibinfo {author} {\bibfnamefont {H.}~\bibnamefont {Heimonen}}, \bibinfo {author} {\bibfnamefont {J.~S.}\ \bibnamefont {Kottmann}}, \bibinfo {author} {\bibfnamefont {T.}~\bibnamefont {Menke}}, \emph {et~al.},\ }\bibfield  {title} {\bibinfo {title} {{Noisy intermediate-scale quantum algorithms}},\ }\href {https://doi.org/10.1103/RevModPhys.94.015004} {\bibfield  {journal} {\bibinfo  {journal} {Rev. Mod. Phys.}\ }\textbf {\bibinfo {volume} {94}},\ \bibinfo {pages} {015004} (\bibinfo {year} {2022})}\BibitemShut {NoStop}%
\bibitem [{\citenamefont {Tilly}\ \emph {et~al.}(2022)\citenamefont {Tilly}, \citenamefont {Chen}, \citenamefont {Cao}, \citenamefont {Picozzi}, \citenamefont {Setia}, \citenamefont {Li}, \citenamefont {Grant}, \citenamefont {Wossnig}, \citenamefont {Rungger}, \citenamefont {Booth} \emph {et~al.}}]{Tilly2022Nov}%
  \BibitemOpen
  \bibfield  {author} {\bibinfo {author} {\bibfnamefont {J.}~\bibnamefont {Tilly}}, \bibinfo {author} {\bibfnamefont {H.}~\bibnamefont {Chen}}, \bibinfo {author} {\bibfnamefont {S.}~\bibnamefont {Cao}}, \bibinfo {author} {\bibfnamefont {D.}~\bibnamefont {Picozzi}}, \bibinfo {author} {\bibfnamefont {K.}~\bibnamefont {Setia}}, \bibinfo {author} {\bibfnamefont {Y.}~\bibnamefont {Li}}, \bibinfo {author} {\bibfnamefont {E.}~\bibnamefont {Grant}}, \bibinfo {author} {\bibfnamefont {L.}~\bibnamefont {Wossnig}}, \bibinfo {author} {\bibfnamefont {I.}~\bibnamefont {Rungger}}, \bibinfo {author} {\bibfnamefont {G.~H.}\ \bibnamefont {Booth}}, \emph {et~al.},\ }\bibfield  {title} {\bibinfo {title} {{The Variational Quantum Eigensolver: A review of methods and best practices}},\ }\href {https://doi.org/10.1016/j.physrep.2022.08.003} {\bibfield  {journal} {\bibinfo  {journal} {Phys. Rep.}\ }\textbf {\bibinfo {volume} {986}},\ \bibinfo {pages} {1} (\bibinfo {year} {2022})}\BibitemShut {NoStop}%
\bibitem [{\citenamefont {Cerezo}\ \emph {et~al.}(2021)\citenamefont {Cerezo}, \citenamefont {Arrasmith}, \citenamefont {Babbush}, \citenamefont {Benjamin}, \citenamefont {Endo}, \citenamefont {Fujii}, \citenamefont {McClean}, \citenamefont {Mitarai}, \citenamefont {Yuan}, \citenamefont {Cincio} \emph {et~al.}}]{Cerezo2021Sep}%
  \BibitemOpen
  \bibfield  {author} {\bibinfo {author} {\bibfnamefont {M.}~\bibnamefont {Cerezo}}, \bibinfo {author} {\bibfnamefont {A.}~\bibnamefont {Arrasmith}}, \bibinfo {author} {\bibfnamefont {R.}~\bibnamefont {Babbush}}, \bibinfo {author} {\bibfnamefont {S.~C.}\ \bibnamefont {Benjamin}}, \bibinfo {author} {\bibfnamefont {S.}~\bibnamefont {Endo}}, \bibinfo {author} {\bibfnamefont {K.}~\bibnamefont {Fujii}}, \bibinfo {author} {\bibfnamefont {J.~R.}\ \bibnamefont {McClean}}, \bibinfo {author} {\bibfnamefont {K.}~\bibnamefont {Mitarai}}, \bibinfo {author} {\bibfnamefont {X.}~\bibnamefont {Yuan}}, \bibinfo {author} {\bibfnamefont {L.}~\bibnamefont {Cincio}}, \emph {et~al.},\ }\bibfield  {title} {\bibinfo {title} {{Variational quantum algorithms}},\ }\href {https://doi.org/10.1038/s42254-021-00348-9} {\bibfield  {journal} {\bibinfo  {journal} {Nat. Rev. Phys.}\ }\textbf {\bibinfo {volume} {3}},\ \bibinfo {pages} {625} (\bibinfo {year} {2021})}\BibitemShut {NoStop}%
\bibitem [{\citenamefont {Peruzzo}\ \emph {et~al.}(2014)\citenamefont {Peruzzo}, \citenamefont {McClean}, \citenamefont {Shadbolt}, \citenamefont {Yung}, \citenamefont {Zhou}, \citenamefont {Love}, \citenamefont {Aspuru-Guzik},\ and\ \citenamefont {O{'}Brien}}]{Peruzzo2014Jul}%
  \BibitemOpen
  \bibfield  {author} {\bibinfo {author} {\bibfnamefont {A.}~\bibnamefont {Peruzzo}}, \bibinfo {author} {\bibfnamefont {J.}~\bibnamefont {McClean}}, \bibinfo {author} {\bibfnamefont {P.}~\bibnamefont {Shadbolt}}, \bibinfo {author} {\bibfnamefont {M.-H.}\ \bibnamefont {Yung}}, \bibinfo {author} {\bibfnamefont {X.-Q.}\ \bibnamefont {Zhou}}, \bibinfo {author} {\bibfnamefont {P.~J.}\ \bibnamefont {Love}}, \bibinfo {author} {\bibfnamefont {A.}~\bibnamefont {Aspuru-Guzik}},\ and\ \bibinfo {author} {\bibfnamefont {J.~L.}\ \bibnamefont {O{'}Brien}},\ }\bibfield  {title} {\bibinfo {title} {{A variational eigenvalue solver on a photonic quantum processor}},\ }\href {https://doi.org/10.1038/ncomms5213} {\bibfield  {journal} {\bibinfo  {journal} {Nat. Commun.}\ }\textbf {\bibinfo {volume} {5}},\ \bibinfo {pages} {1} (\bibinfo {year} {2014})}\BibitemShut {NoStop}%
\bibitem [{\citenamefont {McClean}\ \emph {et~al.}(2016)\citenamefont {McClean}, \citenamefont {Romero}, \citenamefont {Babbush},\ and\ \citenamefont {Aspuru-Guzik}}]{McClean2016Feb}%
  \BibitemOpen
  \bibfield  {author} {\bibinfo {author} {\bibfnamefont {J.~R.}\ \bibnamefont {McClean}}, \bibinfo {author} {\bibfnamefont {J.}~\bibnamefont {Romero}}, \bibinfo {author} {\bibfnamefont {R.}~\bibnamefont {Babbush}},\ and\ \bibinfo {author} {\bibfnamefont {A.}~\bibnamefont {Aspuru-Guzik}},\ }\bibfield  {title} {\bibinfo {title} {{The theory of variational hybrid quantum-classical algorithms}},\ }\href {https://doi.org/10.1088/1367-2630/18/2/023023} {\bibfield  {journal} {\bibinfo  {journal} {New J. Phys.}\ }\textbf {\bibinfo {volume} {18}},\ \bibinfo {pages} {023023} (\bibinfo {year} {2016})}\BibitemShut {NoStop}%
\bibitem [{\citenamefont {Wecker}\ \emph {et~al.}(2015)\citenamefont {Wecker}, \citenamefont {Hastings},\ and\ \citenamefont {Troyer}}]{Wecker2015Oct}%
  \BibitemOpen
  \bibfield  {author} {\bibinfo {author} {\bibfnamefont {D.}~\bibnamefont {Wecker}}, \bibinfo {author} {\bibfnamefont {M.~B.}\ \bibnamefont {Hastings}},\ and\ \bibinfo {author} {\bibfnamefont {M.}~\bibnamefont {Troyer}},\ }\bibfield  {title} {\bibinfo {title} {{Progress towards practical quantum variational algorithms}},\ }\href {https://doi.org/10.1103/PhysRevA.92.042303} {\bibfield  {journal} {\bibinfo  {journal} {Phys. Rev. A}\ }\textbf {\bibinfo {volume} {92}},\ \bibinfo {pages} {042303} (\bibinfo {year} {2015})}\BibitemShut {NoStop}%
\bibitem [{\citenamefont {Motta}\ \emph {et~al.}(2020)\citenamefont {Motta}, \citenamefont {Sun}, \citenamefont {Tan}, \citenamefont {O{'}Rourke}, \citenamefont {Ye}, \citenamefont {Minnich}, \citenamefont {Brand{\ifmmode\tilde{a}\else\~{a}\fi}o},\ and\ \citenamefont {Chan}}]{Motta_2019}%
  \BibitemOpen
  \bibfield  {author} {\bibinfo {author} {\bibfnamefont {M.}~\bibnamefont {Motta}}, \bibinfo {author} {\bibfnamefont {C.}~\bibnamefont {Sun}}, \bibinfo {author} {\bibfnamefont {A.~T.~K.}\ \bibnamefont {Tan}}, \bibinfo {author} {\bibfnamefont {M.~J.}\ \bibnamefont {O{'}Rourke}}, \bibinfo {author} {\bibfnamefont {E.}~\bibnamefont {Ye}}, \bibinfo {author} {\bibfnamefont {A.~J.}\ \bibnamefont {Minnich}}, \bibinfo {author} {\bibfnamefont {F.~G. S.~L.}\ \bibnamefont {Brand{\ifmmode\tilde{a}\else\~{a}\fi}o}},\ and\ \bibinfo {author} {\bibfnamefont {G.~K.-L.}\ \bibnamefont {Chan}},\ }\bibfield  {title} {\bibinfo {title} {{Determining eigenstates and thermal states on a quantum computer using quantum imaginary time evolution}},\ }\href {https://doi.org/10.1038/s41567-019-0704-4} {\bibfield  {journal} {\bibinfo  {journal} {Nat. Phys.}\ }\textbf {\bibinfo {volume} {16}},\ \bibinfo {pages} {205} (\bibinfo {year} {2020})}\BibitemShut {NoStop}%
\bibitem [{\citenamefont {McArdle}\ \emph {et~al.}(2019)\citenamefont {McArdle}, \citenamefont {Jones}, \citenamefont {Endo}, \citenamefont {Li}, \citenamefont {Benjamin},\ and\ \citenamefont {Yuan}}]{McArdle_2019}%
  \BibitemOpen
  \bibfield  {author} {\bibinfo {author} {\bibfnamefont {S.}~\bibnamefont {McArdle}}, \bibinfo {author} {\bibfnamefont {T.}~\bibnamefont {Jones}}, \bibinfo {author} {\bibfnamefont {S.}~\bibnamefont {Endo}}, \bibinfo {author} {\bibfnamefont {Y.}~\bibnamefont {Li}}, \bibinfo {author} {\bibfnamefont {S.~C.}\ \bibnamefont {Benjamin}},\ and\ \bibinfo {author} {\bibfnamefont {X.}~\bibnamefont {Yuan}},\ }\bibfield  {title} {\bibinfo {title} {{Variational ansatz-based quantum simulation of imaginary time evolution}},\ }\href {https://doi.org/10.1038/s41534-019-0187-2} {\bibfield  {journal} {\bibinfo  {journal} {npj Quantum Inf.}\ }\textbf {\bibinfo {volume} {5}},\ \bibinfo {pages} {1} (\bibinfo {year} {2019})}\BibitemShut {NoStop}%
\bibitem [{\citenamefont {Nishi}\ \emph {et~al.}(2021)\citenamefont {Nishi}, \citenamefont {Kosugi},\ and\ \citenamefont {Matsushita}}]{Nishi_2021}%
  \BibitemOpen
  \bibfield  {author} {\bibinfo {author} {\bibfnamefont {H.}~\bibnamefont {Nishi}}, \bibinfo {author} {\bibfnamefont {T.}~\bibnamefont {Kosugi}},\ and\ \bibinfo {author} {\bibfnamefont {Y.-i.}\ \bibnamefont {Matsushita}},\ }\bibfield  {title} {\bibinfo {title} {{Implementation of quantum imaginary-time evolution method on NISQ devices by introducing nonlocal approximation}},\ }\href {https://doi.org/10.1038/s41534-021-00409-y} {\bibfield  {journal} {\bibinfo  {journal} {npj Quantum Inf.}\ }\textbf {\bibinfo {volume} {7}},\ \bibinfo {pages} {1} (\bibinfo {year} {2021})}\BibitemShut {NoStop}%
\bibitem [{\citenamefont {Foulkes}\ \emph {et~al.}(2001)\citenamefont {Foulkes}, \citenamefont {Mitas}, \citenamefont {Needs},\ and\ \citenamefont {Rajagopal}}]{RevModPhys.73.33}%
  \BibitemOpen
  \bibfield  {author} {\bibinfo {author} {\bibfnamefont {W.~M.~C.}\ \bibnamefont {Foulkes}}, \bibinfo {author} {\bibfnamefont {L.}~\bibnamefont {Mitas}}, \bibinfo {author} {\bibfnamefont {R.~J.}\ \bibnamefont {Needs}},\ and\ \bibinfo {author} {\bibfnamefont {G.}~\bibnamefont {Rajagopal}},\ }\bibfield  {title} {\bibinfo {title} {Quantum monte carlo simulations of solids},\ }\href {https://doi.org/10.1103/RevModPhys.73.33} {\bibfield  {journal} {\bibinfo  {journal} {Rev. Mod. Phys.}\ }\textbf {\bibinfo {volume} {73}},\ \bibinfo {pages} {33} (\bibinfo {year} {2001})}\BibitemShut {NoStop}%
\bibitem [{\citenamefont {Yeter-Aydeniz}\ \emph {et~al.}(2020)\citenamefont {Yeter-Aydeniz}, \citenamefont {Pooser},\ and\ \citenamefont {Siopsis}}]{Yeter-Aydeniz2020Jul}%
  \BibitemOpen
  \bibfield  {author} {\bibinfo {author} {\bibfnamefont {K.}~\bibnamefont {Yeter-Aydeniz}}, \bibinfo {author} {\bibfnamefont {R.~C.}\ \bibnamefont {Pooser}},\ and\ \bibinfo {author} {\bibfnamefont {G.}~\bibnamefont {Siopsis}},\ }\bibfield  {title} {\bibinfo {title} {{Practical quantum computation of chemical and nuclear energy levels using quantum imaginary time evolution and Lanczos algorithms}},\ }\href {https://doi.org/10.1038/s41534-020-00290-1} {\bibfield  {journal} {\bibinfo  {journal} {npj Quantum Inf.}\ }\textbf {\bibinfo {volume} {6}},\ \bibinfo {pages} {1} (\bibinfo {year} {2020})}\BibitemShut {NoStop}%
\bibitem [{\citenamefont {Tsuchimochi}\ \emph {et~al.}(2023)\citenamefont {Tsuchimochi}, \citenamefont {Ryo}, \citenamefont {Ten-no},\ and\ \citenamefont {Sasasako}}]{Tsuchimochi2022May}%
  \BibitemOpen
  \bibfield  {author} {\bibinfo {author} {\bibfnamefont {T.}~\bibnamefont {Tsuchimochi}}, \bibinfo {author} {\bibfnamefont {Y.}~\bibnamefont {Ryo}}, \bibinfo {author} {\bibfnamefont {S.~L.}\ \bibnamefont {Ten-no}},\ and\ \bibinfo {author} {\bibfnamefont {K.}~\bibnamefont {Sasasako}},\ }\bibfield  {title} {\bibinfo {title} {Improved algorithms of quantum imaginary time evolution for ground and excited states of molecular systems},\ }\href {https://doi.org/10.1021/acs.jctc.2c00906} {\bibfield  {journal} {\bibinfo  {journal} {Journal of Chemical Theory and Computation}\ }\textbf {\bibinfo {volume} {19}},\ \bibinfo {pages} {503} (\bibinfo {year} {2023})}\BibitemShut {NoStop}%
\bibitem [{\citenamefont {Yuan}\ \emph {et~al.}(2019)\citenamefont {Yuan}, \citenamefont {Endo}, \citenamefont {Zhao}, \citenamefont {Li},\ and\ \citenamefont {Benjamin}}]{Yuan2019Oct}%
  \BibitemOpen
  \bibfield  {author} {\bibinfo {author} {\bibfnamefont {X.}~\bibnamefont {Yuan}}, \bibinfo {author} {\bibfnamefont {S.}~\bibnamefont {Endo}}, \bibinfo {author} {\bibfnamefont {Q.}~\bibnamefont {Zhao}}, \bibinfo {author} {\bibfnamefont {Y.}~\bibnamefont {Li}},\ and\ \bibinfo {author} {\bibfnamefont {S.~C.}\ \bibnamefont {Benjamin}},\ }\bibfield  {title} {\bibinfo {title} {{Theory of variational quantum simulation}},\ }\href {https://doi.org/10.22331/q-2019-10-07-191} {\bibfield  {journal} {\bibinfo  {journal} {Quantum}\ }\textbf {\bibinfo {volume} {3}},\ \bibinfo {pages} {191} (\bibinfo {year} {2019})},\ \Eprint {https://arxiv.org/abs/1812.08767v4} {1812.08767v4} \BibitemShut {NoStop}%
\bibitem [{\citenamefont {Endo}\ \emph {et~al.}(2020)\citenamefont {Endo}, \citenamefont {Sun}, \citenamefont {Li}, \citenamefont {Benjamin},\ and\ \citenamefont {Yuan}}]{PhysRevLett.125.010501}%
  \BibitemOpen
  \bibfield  {author} {\bibinfo {author} {\bibfnamefont {S.}~\bibnamefont {Endo}}, \bibinfo {author} {\bibfnamefont {J.}~\bibnamefont {Sun}}, \bibinfo {author} {\bibfnamefont {Y.}~\bibnamefont {Li}}, \bibinfo {author} {\bibfnamefont {S.~C.}\ \bibnamefont {Benjamin}},\ and\ \bibinfo {author} {\bibfnamefont {X.}~\bibnamefont {Yuan}},\ }\bibfield  {title} {\bibinfo {title} {Variational quantum simulation of general processes},\ }\href {https://doi.org/10.1103/PhysRevLett.125.010501} {\bibfield  {journal} {\bibinfo  {journal} {Phys. Rev. Lett.}\ }\textbf {\bibinfo {volume} {125}},\ \bibinfo {pages} {010501} (\bibinfo {year} {2020})}\BibitemShut {NoStop}%
\bibitem [{\citenamefont {Gomes}\ \emph {et~al.}(2020)\citenamefont {Gomes}, \citenamefont {Zhang}, \citenamefont {Berthusen}, \citenamefont {Wang}, \citenamefont {Ho}, \citenamefont {Orth},\ and\ \citenamefont {Yao}}]{Gomes2020Oct}%
  \BibitemOpen
  \bibfield  {author} {\bibinfo {author} {\bibfnamefont {N.}~\bibnamefont {Gomes}}, \bibinfo {author} {\bibfnamefont {F.}~\bibnamefont {Zhang}}, \bibinfo {author} {\bibfnamefont {N.~F.}\ \bibnamefont {Berthusen}}, \bibinfo {author} {\bibfnamefont {C.-Z.}\ \bibnamefont {Wang}}, \bibinfo {author} {\bibfnamefont {K.-M.}\ \bibnamefont {Ho}}, \bibinfo {author} {\bibfnamefont {P.~P.}\ \bibnamefont {Orth}},\ and\ \bibinfo {author} {\bibfnamefont {Y.}~\bibnamefont {Yao}},\ }\bibfield  {title} {\bibinfo {title} {{Efficient Step-Merged Quantum Imaginary Time Evolution Algorithm for Quantum Chemistry}},\ }\href {https://doi.org/10.1021/acs.jctc.0c00666} {\bibfield  {journal} {\bibinfo  {journal} {J. Chem. Theory Comput.}\ }\textbf {\bibinfo {volume} {16}},\ \bibinfo {pages} {6256} (\bibinfo {year} {2020})}\BibitemShut {NoStop}%
\bibitem [{\citenamefont {Wick}(1954)}]{PhysRev.96.1124}%
  \BibitemOpen
  \bibfield  {author} {\bibinfo {author} {\bibfnamefont {G.~C.}\ \bibnamefont {Wick}},\ }\bibfield  {title} {\bibinfo {title} {Properties of bethe-salpeter wave functions},\ }\href {https://doi.org/10.1103/PhysRev.96.1124} {\bibfield  {journal} {\bibinfo  {journal} {Phys. Rev.}\ }\textbf {\bibinfo {volume} {96}},\ \bibinfo {pages} {1124} (\bibinfo {year} {1954})}\BibitemShut {NoStop}%
\bibitem [{\citenamefont {Lee}\ \emph {et~al.}(2023)\citenamefont {Lee}, \citenamefont {Lee}, \citenamefont {Zhai}, \citenamefont {Tong}, \citenamefont {Dalzell}, \citenamefont {Kumar}, \citenamefont {Helms}, \citenamefont {Gray}, \citenamefont {Cui}, \citenamefont {Liu}, \citenamefont {Kastoryano}, \citenamefont {Babbush}, \citenamefont {Preskill}, \citenamefont {Reichman}, \citenamefont {Campbell}, \citenamefont {Valeev}, \citenamefont {Lin},\ and\ \citenamefont {Chan}}]{Lee2022Aug}%
  \BibitemOpen
  \bibfield  {author} {\bibinfo {author} {\bibfnamefont {S.}~\bibnamefont {Lee}}, \bibinfo {author} {\bibfnamefont {J.}~\bibnamefont {Lee}}, \bibinfo {author} {\bibfnamefont {H.}~\bibnamefont {Zhai}}, \bibinfo {author} {\bibfnamefont {Y.}~\bibnamefont {Tong}}, \bibinfo {author} {\bibfnamefont {A.~M.}\ \bibnamefont {Dalzell}}, \bibinfo {author} {\bibfnamefont {A.}~\bibnamefont {Kumar}}, \bibinfo {author} {\bibfnamefont {P.}~\bibnamefont {Helms}}, \bibinfo {author} {\bibfnamefont {J.}~\bibnamefont {Gray}}, \bibinfo {author} {\bibfnamefont {Z.-H.}\ \bibnamefont {Cui}}, \bibinfo {author} {\bibfnamefont {W.}~\bibnamefont {Liu}}, \bibinfo {author} {\bibfnamefont {M.}~\bibnamefont {Kastoryano}}, \bibinfo {author} {\bibfnamefont {R.}~\bibnamefont {Babbush}}, \bibinfo {author} {\bibfnamefont {J.}~\bibnamefont {Preskill}}, \bibinfo {author} {\bibfnamefont {D.~R.}\ \bibnamefont {Reichman}}, \bibinfo {author} {\bibfnamefont {E.~T.}\ \bibnamefont {Campbell}}, \bibinfo {author} {\bibfnamefont {E.~F.}\ \bibnamefont
  {Valeev}}, \bibinfo {author} {\bibfnamefont {L.}~\bibnamefont {Lin}},\ and\ \bibinfo {author} {\bibfnamefont {G.~K.-L.}\ \bibnamefont {Chan}},\ }\bibfield  {title} {\bibinfo {title} {Evaluating the evidence for exponential quantum advantage in ground-state quantum chemistry},\ }\href {https://doi.org/10.1038/s41467-023-37587-6} {\bibfield  {journal} {\bibinfo  {journal} {Nature Communications}\ }\textbf {\bibinfo {volume} {14}},\ \bibinfo {pages} {1952} (\bibinfo {year} {2023})}\BibitemShut {NoStop}%
\bibitem [{\citenamefont {Kandala}\ \emph {et~al.}(2017)\citenamefont {Kandala}, \citenamefont {Mezzacapo}, \citenamefont {Temme}, \citenamefont {Takita}, \citenamefont {Brink}, \citenamefont {Chow},\ and\ \citenamefont {Gambetta}}]{Kandala2017Sep}%
  \BibitemOpen
  \bibfield  {author} {\bibinfo {author} {\bibfnamefont {A.}~\bibnamefont {Kandala}}, \bibinfo {author} {\bibfnamefont {A.}~\bibnamefont {Mezzacapo}}, \bibinfo {author} {\bibfnamefont {K.}~\bibnamefont {Temme}}, \bibinfo {author} {\bibfnamefont {M.}~\bibnamefont {Takita}}, \bibinfo {author} {\bibfnamefont {M.}~\bibnamefont {Brink}}, \bibinfo {author} {\bibfnamefont {J.~M.}\ \bibnamefont {Chow}},\ and\ \bibinfo {author} {\bibfnamefont {J.~M.}\ \bibnamefont {Gambetta}},\ }\bibfield  {title} {\bibinfo {title} {{Hardware-efficient variational quantum eigensolver for small molecules and quantum magnets}},\ }\href {https://doi.org/10.1038/nature23879} {\bibfield  {journal} {\bibinfo  {journal} {Nature}\ }\textbf {\bibinfo {volume} {549}},\ \bibinfo {pages} {242} (\bibinfo {year} {2017})}\BibitemShut {NoStop}%
\bibitem [{\citenamefont {Kyaw}\ \emph {et~al.}(2023)\citenamefont {Kyaw}, \citenamefont {Soley}, \citenamefont {Allen}, \citenamefont {Bergold}, \citenamefont {Sun}, \citenamefont {Batista},\ and\ \citenamefont {Aspuru-Guzik}}]{Kyaw2022Aug}%
  \BibitemOpen
  \bibfield  {author} {\bibinfo {author} {\bibfnamefont {T.~H.}\ \bibnamefont {Kyaw}}, \bibinfo {author} {\bibfnamefont {M.~B.}\ \bibnamefont {Soley}}, \bibinfo {author} {\bibfnamefont {B.}~\bibnamefont {Allen}}, \bibinfo {author} {\bibfnamefont {P.}~\bibnamefont {Bergold}}, \bibinfo {author} {\bibfnamefont {C.}~\bibnamefont {Sun}}, \bibinfo {author} {\bibfnamefont {V.~S.}\ \bibnamefont {Batista}},\ and\ \bibinfo {author} {\bibfnamefont {A.}~\bibnamefont {Aspuru-Guzik}},\ }\bibfield  {title} {\bibinfo {title} {{Boosting quantum amplitude exponentially in variational quantum algorithms}},\ }\href {https://doi.org/10.1088/2058-9565/acf4ba} {\bibfield  {journal} {\bibinfo  {journal} {Quantum Sci. Technol.}\ }\textbf {\bibinfo {volume} {9}},\ \bibinfo {pages} {01LT01} (\bibinfo {year} {2023})}\BibitemShut {NoStop}%
\bibitem [{\citenamefont {Zhang}\ \emph {et~al.}(2023)\citenamefont {Zhang}, \citenamefont {Sun}, \citenamefont {Yuan},\ and\ \citenamefont {Yung}}]{Zhang2023Jan}%
  \BibitemOpen
  \bibfield  {author} {\bibinfo {author} {\bibfnamefont {Z.-J.}\ \bibnamefont {Zhang}}, \bibinfo {author} {\bibfnamefont {J.}~\bibnamefont {Sun}}, \bibinfo {author} {\bibfnamefont {X.}~\bibnamefont {Yuan}},\ and\ \bibinfo {author} {\bibfnamefont {M.-H.}\ \bibnamefont {Yung}},\ }\bibfield  {title} {\bibinfo {title} {{Low-Depth Hamiltonian Simulation by an Adaptive Product Formula}},\ }\href {https://doi.org/10.1103/PhysRevLett.130.040601} {\bibfield  {journal} {\bibinfo  {journal} {Phys. Rev. Lett.}\ }\textbf {\bibinfo {volume} {130}},\ \bibinfo {pages} {040601} (\bibinfo {year} {2023})}\BibitemShut {NoStop}%
\bibitem [{\citenamefont {Benedetti}\ \emph {et~al.}(2021)\citenamefont {Benedetti}, \citenamefont {Fiorentini},\ and\ \citenamefont {Lubasch}}]{PhysRevResearch.3.033083}%
  \BibitemOpen
  \bibfield  {author} {\bibinfo {author} {\bibfnamefont {M.}~\bibnamefont {Benedetti}}, \bibinfo {author} {\bibfnamefont {M.}~\bibnamefont {Fiorentini}},\ and\ \bibinfo {author} {\bibfnamefont {M.}~\bibnamefont {Lubasch}},\ }\bibfield  {title} {\bibinfo {title} {Hardware-efficient variational quantum algorithms for time evolution},\ }\href {https://doi.org/10.1103/PhysRevResearch.3.033083} {\bibfield  {journal} {\bibinfo  {journal} {Phys. Rev. Res.}\ }\textbf {\bibinfo {volume} {3}},\ \bibinfo {pages} {033083} (\bibinfo {year} {2021})}\BibitemShut {NoStop}%
\bibitem [{\citenamefont {Zoufal}\ \emph {et~al.}(2023)\citenamefont {Zoufal}, \citenamefont {Sutter},\ and\ \citenamefont {Woerner}}]{zoufal2021error}%
  \BibitemOpen
  \bibfield  {author} {\bibinfo {author} {\bibfnamefont {C.}~\bibnamefont {Zoufal}}, \bibinfo {author} {\bibfnamefont {D.}~\bibnamefont {Sutter}},\ and\ \bibinfo {author} {\bibfnamefont {S.}~\bibnamefont {Woerner}},\ }\bibfield  {title} {\bibinfo {title} {Error bounds for variational quantum time evolution},\ }\href {https://doi.org/10.1103/PhysRevApplied.20.044059} {\bibfield  {journal} {\bibinfo  {journal} {Phys. Rev. Appl.}\ }\textbf {\bibinfo {volume} {20}},\ \bibinfo {pages} {044059} (\bibinfo {year} {2023})}\BibitemShut {NoStop}%
\bibitem [{\citenamefont {Stilck~Fran{\ifmmode\mbox{\c{c}}\else\c{c}\fi}a}\ and\ \citenamefont {Garc{\ifmmode\acute{\imath}\else\'{\i}\fi}a-Patr{\ifmmode\acute{o}\else\'{o}\fi}n}(2021)}]{StilckFranca2021Nov}%
  \BibitemOpen
  \bibfield  {author} {\bibinfo {author} {\bibfnamefont {D.}~\bibnamefont {Stilck~Fran{\ifmmode\mbox{\c{c}}\else\c{c}\fi}a}}\ and\ \bibinfo {author} {\bibfnamefont {R.}~\bibnamefont {Garc{\ifmmode\acute{\imath}\else\'{\i}\fi}a-Patr{\ifmmode\acute{o}\else\'{o}\fi}n}},\ }\bibfield  {title} {\bibinfo {title} {{Limitations of optimization algorithms on noisy quantum devices}},\ }\href {https://doi.org/10.1038/s41567-021-01356-3} {\bibfield  {journal} {\bibinfo  {journal} {Nat. Phys.}\ }\textbf {\bibinfo {volume} {17}},\ \bibinfo {pages} {1221} (\bibinfo {year} {2021})}\BibitemShut {NoStop}%
\bibitem [{\citenamefont {Wang}\ \emph {et~al.}(2021)\citenamefont {Wang}, \citenamefont {Fontana}, \citenamefont {Cerezo}, \citenamefont {Sharma}, \citenamefont {Sone}, \citenamefont {Cincio},\ and\ \citenamefont {Coles}}]{Wang2021Nov}%
  \BibitemOpen
  \bibfield  {author} {\bibinfo {author} {\bibfnamefont {S.}~\bibnamefont {Wang}}, \bibinfo {author} {\bibfnamefont {E.}~\bibnamefont {Fontana}}, \bibinfo {author} {\bibfnamefont {M.}~\bibnamefont {Cerezo}}, \bibinfo {author} {\bibfnamefont {K.}~\bibnamefont {Sharma}}, \bibinfo {author} {\bibfnamefont {A.}~\bibnamefont {Sone}}, \bibinfo {author} {\bibfnamefont {L.}~\bibnamefont {Cincio}},\ and\ \bibinfo {author} {\bibfnamefont {P.~J.}\ \bibnamefont {Coles}},\ }\bibfield  {title} {\bibinfo {title} {{Noise-induced barren plateaus in variational quantum algorithms}},\ }\href {https://doi.org/10.1038/s41467-021-27045-6} {\bibfield  {journal} {\bibinfo  {journal} {Nat. Commun.}\ }\textbf {\bibinfo {volume} {12}},\ \bibinfo {pages} {1} (\bibinfo {year} {2021})}\BibitemShut {NoStop}%
\bibitem [{Note1()}]{Note1}%
  \BibitemOpen
  \bibinfo {note} {Note two minor corrections in Fig.~3 of Ref.~\cite {Yuan2019Oct}. First one is that $U_k:= S_k^{\dagger }T_k$, and the second is that in the second system, $T_q$ will come after $U_q$ and not $S_q$ (as given in Fig.~3 of \cite {Yuan2019Oct}) as the definition of $M_{k,q}$ involves taking the conjugate of one term.}\BibitemShut {Stop}%
\bibitem [{\citenamefont {Xue}\ \emph {et~al.}(2021)\citenamefont {Xue}, \citenamefont {Chen}, \citenamefont {Wu},\ and\ \citenamefont {Guo}}]{Xue2021Mar}%
  \BibitemOpen
  \bibfield  {author} {\bibinfo {author} {\bibfnamefont {C.}~\bibnamefont {Xue}}, \bibinfo {author} {\bibfnamefont {Z.-Y.}\ \bibnamefont {Chen}}, \bibinfo {author} {\bibfnamefont {Y.-C.}\ \bibnamefont {Wu}},\ and\ \bibinfo {author} {\bibfnamefont {G.-P.}\ \bibnamefont {Guo}},\ }\bibfield  {title} {\bibinfo {title} {{Effects of Quantum Noise on Quantum Approximate Optimization Algorithm}},\ }\href {https://doi.org/10.1088/0256-307X/38/3/030302} {\bibfield  {journal} {\bibinfo  {journal} {Chin. Phys. Lett.}\ }\textbf {\bibinfo {volume} {38}},\ \bibinfo {pages} {030302} (\bibinfo {year} {2021})}\BibitemShut {NoStop}%
\bibitem [{\citenamefont {Marshall}\ \emph {et~al.}(2020)\citenamefont {Marshall}, \citenamefont {Wudarski}, \citenamefont {Hadfield},\ and\ \citenamefont {Hogg}}]{Marshall2020Aug}%
  \BibitemOpen
  \bibfield  {author} {\bibinfo {author} {\bibfnamefont {J.}~\bibnamefont {Marshall}}, \bibinfo {author} {\bibfnamefont {F.}~\bibnamefont {Wudarski}}, \bibinfo {author} {\bibfnamefont {S.}~\bibnamefont {Hadfield}},\ and\ \bibinfo {author} {\bibfnamefont {T.}~\bibnamefont {Hogg}},\ }\bibfield  {title} {\bibinfo {title} {{Characterizing local noise in QAOA circuits}},\ }\href {https://doi.org/10.1088/2633-1357/abb0d7} {\bibfield  {journal} {\bibinfo  {journal} {IOP SciNotes}\ }\textbf {\bibinfo {volume} {1}},\ \bibinfo {pages} {025208} (\bibinfo {year} {2020})}\BibitemShut {NoStop}%
\bibitem [{Note2()}]{Note2}%
  \BibitemOpen
  \bibinfo {note} {We denote $\|\cdot \|_2$ and $\|\cdot \|$ interchangeably to indicate the Frobenius norm defined as $\|A\|_2 = \protect \sqrt {{\protect \rm Tr}(A^* A)}$}\BibitemShut {NoStop}%
\bibitem [{Note3()}]{Note3}%
  \BibitemOpen
  \bibinfo {note} {We denote a system and its corresponding Hilbert space using an uppercase letter with a numerical subscript, like $A_0$ or $A_1$.}\BibitemShut {Stop}%
\bibitem [{\citenamefont {Bastidas}\ \emph {et~al.}(2018)\citenamefont {Bastidas}, \citenamefont {Kyaw}, \citenamefont {Tangpanitanon}, \citenamefont {Romero}, \citenamefont {Kwek},\ and\ \citenamefont {Angelakis}}]{Bastidas2018Sep}%
  \BibitemOpen
  \bibfield  {author} {\bibinfo {author} {\bibfnamefont {V.~M.}\ \bibnamefont {Bastidas}}, \bibinfo {author} {\bibfnamefont {T.~H.}\ \bibnamefont {Kyaw}}, \bibinfo {author} {\bibfnamefont {J.}~\bibnamefont {Tangpanitanon}}, \bibinfo {author} {\bibfnamefont {G.}~\bibnamefont {Romero}}, \bibinfo {author} {\bibfnamefont {L.-C.}\ \bibnamefont {Kwek}},\ and\ \bibinfo {author} {\bibfnamefont {D.~G.}\ \bibnamefont {Angelakis}},\ }\bibfield  {title} {\bibinfo {title} {{Floquet stroboscopic divisibility in non-Markovian dynamics}},\ }\href {https://doi.org/10.1088/1367-2630/aadcbd} {\bibfield  {journal} {\bibinfo  {journal} {New J. Phys.}\ }\textbf {\bibinfo {volume} {20}},\ \bibinfo {pages} {093004} (\bibinfo {year} {2018})}\BibitemShut {NoStop}%
\bibitem [{\citenamefont {Kyaw}\ \emph {et~al.}(2017)\citenamefont {Kyaw}, \citenamefont {Allende}, \citenamefont {Kwek},\ and\ \citenamefont {Romero}}]{Kyaw2017May}%
  \BibitemOpen
  \bibfield  {author} {\bibinfo {author} {\bibfnamefont {T.~H.}\ \bibnamefont {Kyaw}}, \bibinfo {author} {\bibfnamefont {S.}~\bibnamefont {Allende}}, \bibinfo {author} {\bibfnamefont {L.-C.}\ \bibnamefont {Kwek}},\ and\ \bibinfo {author} {\bibfnamefont {G.}~\bibnamefont {Romero}},\ }\bibfield  {title} {\bibinfo {title} {{Parity-preserving light-matter system mediates effective two-body interactions}},\ }\href {https://doi.org/10.1088/2058-9565/aa701c} {\bibfield  {journal} {\bibinfo  {journal} {Quantum Sci. Technol.}\ }\textbf {\bibinfo {volume} {2}},\ \bibinfo {pages} {025007} (\bibinfo {year} {2017})}\BibitemShut {NoStop}%
\bibitem [{\citenamefont {Gour}(2019)}]{Gour2019Mar}%
  \BibitemOpen
  \bibfield  {author} {\bibinfo {author} {\bibfnamefont {G.}~\bibnamefont {Gour}},\ }\bibfield  {title} {\bibinfo {title} {{Comparison of Quantum Channels by Superchannels}},\ }\href {https://doi.org/10.1109/TIT.2019.2907989} {\bibfield  {journal} {\bibinfo  {journal} {IEEE Trans. Inf. Theory}\ }\textbf {\bibinfo {volume} {65}},\ \bibinfo {pages} {5880} (\bibinfo {year} {2019})}\BibitemShut {NoStop}%
\bibitem [{\citenamefont {Regula}\ and\ \citenamefont {Takagi}(2021)}]{Regula2021Jul}%
  \BibitemOpen
  \bibfield  {author} {\bibinfo {author} {\bibfnamefont {B.}~\bibnamefont {Regula}}\ and\ \bibinfo {author} {\bibfnamefont {R.}~\bibnamefont {Takagi}},\ }\bibfield  {title} {\bibinfo {title} {{Fundamental limitations on distillation of quantum channel resources}},\ }\href {https://doi.org/10.1038/s41467-021-24699-0} {\bibfield  {journal} {\bibinfo  {journal} {Nat. Commun.}\ }\textbf {\bibinfo {volume} {12}},\ \bibinfo {pages} {1} (\bibinfo {year} {2021})}\BibitemShut {NoStop}%
\bibitem [{\citenamefont {Mi}\ \emph {et~al.}(2022)\citenamefont {Mi}, \citenamefont {Sonner}, \citenamefont {Niu}, \citenamefont {Lee}, \citenamefont {Foxen}, \citenamefont {Acharya}, \citenamefont {Aleiner}, \citenamefont {Andersen}, \citenamefont {Arute}, \citenamefont {Arya} \emph {et~al.}}]{Mi2022Nov}%
  \BibitemOpen
  \bibfield  {author} {\bibinfo {author} {\bibfnamefont {X.}~\bibnamefont {Mi}}, \bibinfo {author} {\bibfnamefont {M.}~\bibnamefont {Sonner}}, \bibinfo {author} {\bibfnamefont {M.~Y.}\ \bibnamefont {Niu}}, \bibinfo {author} {\bibfnamefont {K.~W.}\ \bibnamefont {Lee}}, \bibinfo {author} {\bibfnamefont {B.}~\bibnamefont {Foxen}}, \bibinfo {author} {\bibfnamefont {R.}~\bibnamefont {Acharya}}, \bibinfo {author} {\bibfnamefont {I.}~\bibnamefont {Aleiner}}, \bibinfo {author} {\bibfnamefont {T.~I.}\ \bibnamefont {Andersen}}, \bibinfo {author} {\bibfnamefont {F.}~\bibnamefont {Arute}}, \bibinfo {author} {\bibfnamefont {K.}~\bibnamefont {Arya}}, \emph {et~al.},\ }\bibfield  {title} {\bibinfo {title} {Noise-resilient edge modes on a chain of superconducting qubits},\ }\href {https://doi.org/10.1126/science.abq5769} {\bibfield  {journal} {\bibinfo  {journal} {Science}\ }\textbf {\bibinfo {volume} {378}},\ \bibinfo {pages} {785} (\bibinfo {year} {2022})}\BibitemShut {NoStop}%
\bibitem [{\citenamefont {Georgopoulos}\ \emph {et~al.}(2021)\citenamefont {Georgopoulos}, \citenamefont {Emary},\ and\ \citenamefont {Zuliani}}]{PhysRevA.104.062432}%
  \BibitemOpen
  \bibfield  {author} {\bibinfo {author} {\bibfnamefont {K.}~\bibnamefont {Georgopoulos}}, \bibinfo {author} {\bibfnamefont {C.}~\bibnamefont {Emary}},\ and\ \bibinfo {author} {\bibfnamefont {P.}~\bibnamefont {Zuliani}},\ }\bibfield  {title} {\bibinfo {title} {Modeling and simulating the noisy behavior of near-term quantum computers},\ }\href {https://doi.org/10.1103/PhysRevA.104.062432} {\bibfield  {journal} {\bibinfo  {journal} {Phys. Rev. A}\ }\textbf {\bibinfo {volume} {104}},\ \bibinfo {pages} {062432} (\bibinfo {year} {2021})}\BibitemShut {NoStop}%
\bibitem [{\citenamefont {von Petersdorff}(2018)}]{tobias_lect}%
  \BibitemOpen
  \bibfield  {author} {\bibinfo {author} {\bibfnamefont {T.}~\bibnamefont {von Petersdorff}},\ }\href@noop {} {\bibinfo {title} {Errors for linear systems}},\ \bibinfo {howpublished} {\url{https://www.math.umd.edu/~petersd/466/linsysterrn.pdf/}} (\bibinfo {year} {2018})\BibitemShut {NoStop}%
\bibitem [{\citenamefont {Cai}\ \emph {et~al.}(2023)\citenamefont {Cai}, \citenamefont {Babbush}, \citenamefont {Benjamin}, \citenamefont {Endo}, \citenamefont {Huggins}, \citenamefont {Li}, \citenamefont {McClean},\ and\ \citenamefont {O'Brien}}]{Cai2022Oct}%
  \BibitemOpen
  \bibfield  {author} {\bibinfo {author} {\bibfnamefont {Z.}~\bibnamefont {Cai}}, \bibinfo {author} {\bibfnamefont {R.}~\bibnamefont {Babbush}}, \bibinfo {author} {\bibfnamefont {S.~C.}\ \bibnamefont {Benjamin}}, \bibinfo {author} {\bibfnamefont {S.}~\bibnamefont {Endo}}, \bibinfo {author} {\bibfnamefont {W.~J.}\ \bibnamefont {Huggins}}, \bibinfo {author} {\bibfnamefont {Y.}~\bibnamefont {Li}}, \bibinfo {author} {\bibfnamefont {J.~R.}\ \bibnamefont {McClean}},\ and\ \bibinfo {author} {\bibfnamefont {T.~E.}\ \bibnamefont {O'Brien}},\ }\bibfield  {title} {\bibinfo {title} {Quantum error mitigation},\ }\href {https://doi.org/10.1103/RevModPhys.95.045005} {\bibfield  {journal} {\bibinfo  {journal} {Rev. Mod. Phys.}\ }\textbf {\bibinfo {volume} {95}},\ \bibinfo {pages} {045005} (\bibinfo {year} {2023})}\BibitemShut {NoStop}%
\bibitem [{\citenamefont {Quek}\ \emph {et~al.}(2022)\citenamefont {Quek}, \citenamefont {Fran{\ifmmode\mbox{\c{c}}\else\c{c}\fi}a}, \citenamefont {Khatri}, \citenamefont {Meyer},\ and\ \citenamefont {Eisert}}]{Quek2022Oct}%
  \BibitemOpen
  \bibfield  {author} {\bibinfo {author} {\bibfnamefont {Y.}~\bibnamefont {Quek}}, \bibinfo {author} {\bibfnamefont {D.~S.}\ \bibnamefont {Fran{\ifmmode\mbox{\c{c}}\else\c{c}\fi}a}}, \bibinfo {author} {\bibfnamefont {S.}~\bibnamefont {Khatri}}, \bibinfo {author} {\bibfnamefont {J.~J.}\ \bibnamefont {Meyer}},\ and\ \bibinfo {author} {\bibfnamefont {J.}~\bibnamefont {Eisert}},\ }\bibfield  {title} {\bibinfo {title} {{Exponentially tighter bounds on limitations of quantum error mitigation}},\ }\bibfield  {journal} {\bibinfo  {journal} {arXiv}\ }\href {https://doi.org/10.48550/arXiv.2210.11505} {10.48550/arXiv.2210.11505} (\bibinfo {year} {2022}),\ \Eprint {https://arxiv.org/abs/2210.11505} {2210.11505} \BibitemShut {NoStop}%
\bibitem [{\citenamefont {Temme}\ \emph {et~al.}(2017)\citenamefont {Temme}, \citenamefont {Bravyi},\ and\ \citenamefont {Gambetta}}]{Temme2017Nov}%
  \BibitemOpen
  \bibfield  {author} {\bibinfo {author} {\bibfnamefont {K.}~\bibnamefont {Temme}}, \bibinfo {author} {\bibfnamefont {S.}~\bibnamefont {Bravyi}},\ and\ \bibinfo {author} {\bibfnamefont {J.~M.}\ \bibnamefont {Gambetta}},\ }\bibfield  {title} {\bibinfo {title} {{Error Mitigation for Short-Depth Quantum Circuits}},\ }\href {https://doi.org/10.1103/PhysRevLett.119.180509} {\bibfield  {journal} {\bibinfo  {journal} {Phys. Rev. Lett.}\ }\textbf {\bibinfo {volume} {119}},\ \bibinfo {pages} {180509} (\bibinfo {year} {2017})}\BibitemShut {NoStop}%
\bibitem [{\citenamefont {Li}\ and\ \citenamefont {Benjamin}(2017)}]{Li2017Jun}%
  \BibitemOpen
  \bibfield  {author} {\bibinfo {author} {\bibfnamefont {Y.}~\bibnamefont {Li}}\ and\ \bibinfo {author} {\bibfnamefont {S.~C.}\ \bibnamefont {Benjamin}},\ }\bibfield  {title} {\bibinfo {title} {{Efficient Variational Quantum Simulator Incorporating Active Error Minimization}},\ }\href {https://doi.org/10.1103/PhysRevX.7.021050} {\bibfield  {journal} {\bibinfo  {journal} {Phys. Rev. X}\ }\textbf {\bibinfo {volume} {7}},\ \bibinfo {pages} {021050} (\bibinfo {year} {2017})}\BibitemShut {NoStop}%
\bibitem [{\citenamefont {Kandala}\ \emph {et~al.}(2019)\citenamefont {Kandala}, \citenamefont {Temme}, \citenamefont {C{\ifmmode\acute{o}\else\'{o}\fi}rcoles}, \citenamefont {Mezzacapo}, \citenamefont {Chow},\ and\ \citenamefont {Gambetta}}]{Kandala2019Mar}%
  \BibitemOpen
  \bibfield  {author} {\bibinfo {author} {\bibfnamefont {A.}~\bibnamefont {Kandala}}, \bibinfo {author} {\bibfnamefont {K.}~\bibnamefont {Temme}}, \bibinfo {author} {\bibfnamefont {A.~D.}\ \bibnamefont {C{\ifmmode\acute{o}\else\'{o}\fi}rcoles}}, \bibinfo {author} {\bibfnamefont {A.}~\bibnamefont {Mezzacapo}}, \bibinfo {author} {\bibfnamefont {J.~M.}\ \bibnamefont {Chow}},\ and\ \bibinfo {author} {\bibfnamefont {J.~M.}\ \bibnamefont {Gambetta}},\ }\bibfield  {title} {\bibinfo {title} {{Error mitigation extends the computational reach of a noisy quantum processor}},\ }\href {https://doi.org/10.1038/s41586-019-1040-7} {\bibfield  {journal} {\bibinfo  {journal} {Nature}\ }\textbf {\bibinfo {volume} {567}},\ \bibinfo {pages} {491} (\bibinfo {year} {2019})}\BibitemShut {NoStop}%
\bibitem [{\citenamefont {Giurgica-Tiron}\ \emph {et~al.}(2020)\citenamefont {Giurgica-Tiron}, \citenamefont {Hindy}, \citenamefont {LaRose}, \citenamefont {Mari},\ and\ \citenamefont {Zeng}}]{Giurgica-Tiron2020Oct}%
  \BibitemOpen
  \bibfield  {author} {\bibinfo {author} {\bibfnamefont {T.}~\bibnamefont {Giurgica-Tiron}}, \bibinfo {author} {\bibfnamefont {Y.}~\bibnamefont {Hindy}}, \bibinfo {author} {\bibfnamefont {R.}~\bibnamefont {LaRose}}, \bibinfo {author} {\bibfnamefont {A.}~\bibnamefont {Mari}},\ and\ \bibinfo {author} {\bibfnamefont {W.~J.}\ \bibnamefont {Zeng}},\ }\bibfield  {title} {\bibinfo {title} {{Digital zero noise extrapolation for quantum error mitigation}},\ }in\ \href {https://doi.org/10.1109/QCE49297.2020.00045} {\emph {\bibinfo {booktitle} {{2020 IEEE International Conference on Quantum Computing and Engineering (QCE)}}}}\ (\bibinfo  {publisher} {IEEE},\ \bibinfo {address} {Denver, CO, USA},\ \bibinfo {year} {2020})\ pp.\ \bibinfo {pages} {306--316}\BibitemShut {NoStop}%
\bibitem [{\citenamefont {Endo}\ \emph {et~al.}(2018)\citenamefont {Endo}, \citenamefont {Benjamin},\ and\ \citenamefont {Li}}]{Endo2018Jul}%
  \BibitemOpen
  \bibfield  {author} {\bibinfo {author} {\bibfnamefont {S.}~\bibnamefont {Endo}}, \bibinfo {author} {\bibfnamefont {S.~C.}\ \bibnamefont {Benjamin}},\ and\ \bibinfo {author} {\bibfnamefont {Y.}~\bibnamefont {Li}},\ }\bibfield  {title} {\bibinfo {title} {{Practical Quantum Error Mitigation for Near-Future Applications}},\ }\href {https://doi.org/10.1103/PhysRevX.8.031027} {\bibfield  {journal} {\bibinfo  {journal} {Phys. Rev. X}\ }\textbf {\bibinfo {volume} {8}},\ \bibinfo {pages} {031027} (\bibinfo {year} {2018})}\BibitemShut {NoStop}%
\bibitem [{\citenamefont {Song}\ \emph {et~al.}(2019)\citenamefont {Song}, \citenamefont {Cui}, \citenamefont {Wang}, \citenamefont {Hao}, \citenamefont {Feng},\ and\ \citenamefont {Li}}]{Song2019Sep}%
  \BibitemOpen
  \bibfield  {author} {\bibinfo {author} {\bibfnamefont {C.}~\bibnamefont {Song}}, \bibinfo {author} {\bibfnamefont {J.}~\bibnamefont {Cui}}, \bibinfo {author} {\bibfnamefont {H.}~\bibnamefont {Wang}}, \bibinfo {author} {\bibfnamefont {J.}~\bibnamefont {Hao}}, \bibinfo {author} {\bibfnamefont {H.}~\bibnamefont {Feng}},\ and\ \bibinfo {author} {\bibfnamefont {Y.}~\bibnamefont {Li}},\ }\bibfield  {title} {\bibinfo {title} {{Quantum computation with universal error mitigation on a superconducting quantum processor}},\ }\href {https://doi.org/10.1126/sciadv.aaw5686} {\bibfield  {journal} {\bibinfo  {journal} {Sci. Adv.}\ }\textbf {\bibinfo {volume} {5}},\ \bibinfo {pages} {eaaw5686} (\bibinfo {year} {2019})}\BibitemShut {NoStop}%
\bibitem [{\citenamefont {Zhang}\ \emph {et~al.}(2020)\citenamefont {Zhang}, \citenamefont {Lu}, \citenamefont {Zhang}, \citenamefont {Chen}, \citenamefont {Li}, \citenamefont {Zhang},\ and\ \citenamefont {Kim}}]{Zhang2020Jan}%
  \BibitemOpen
  \bibfield  {author} {\bibinfo {author} {\bibfnamefont {S.}~\bibnamefont {Zhang}}, \bibinfo {author} {\bibfnamefont {Y.}~\bibnamefont {Lu}}, \bibinfo {author} {\bibfnamefont {K.}~\bibnamefont {Zhang}}, \bibinfo {author} {\bibfnamefont {W.}~\bibnamefont {Chen}}, \bibinfo {author} {\bibfnamefont {Y.}~\bibnamefont {Li}}, \bibinfo {author} {\bibfnamefont {J.-N.}\ \bibnamefont {Zhang}},\ and\ \bibinfo {author} {\bibfnamefont {K.}~\bibnamefont {Kim}},\ }\bibfield  {title} {\bibinfo {title} {{Error-mitigated quantum gates exceeding physical fidelities in a trapped-ion system}},\ }\href {https://doi.org/10.1038/s41467-020-14376-z} {\bibfield  {journal} {\bibinfo  {journal} {Nat. Commun.}\ }\textbf {\bibinfo {volume} {11}},\ \bibinfo {pages} {1} (\bibinfo {year} {2020})}\BibitemShut {NoStop}%
\bibitem [{\citenamefont {Higham}(2002)}]{Higham2002}%
  \BibitemOpen
  \bibfield  {author} {\bibinfo {author} {\bibfnamefont {N.~J.}\ \bibnamefont {Higham}},\ }\href {https://doi.org/10.1137/1.9780898718027} {\emph {\bibinfo {title} {Accuracy and Stability of Numerical Algorithms}}},\ \bibinfo {edition} {2nd}\ ed.\ (\bibinfo  {publisher} {Society for Industrial and Applied Mathematics},\ \bibinfo {address} {Philadelphia, PA, USA},\ \bibinfo {year} {2002})\BibitemShut {NoStop}%
\bibitem [{\citenamefont {Sim}\ \emph {et~al.}(2019)\citenamefont {Sim}, \citenamefont {Johnson},\ and\ \citenamefont {Aspuru-Guzik}}]{expressibility}%
  \BibitemOpen
  \bibfield  {author} {\bibinfo {author} {\bibfnamefont {S.}~\bibnamefont {Sim}}, \bibinfo {author} {\bibfnamefont {P.~D.}\ \bibnamefont {Johnson}},\ and\ \bibinfo {author} {\bibfnamefont {A.}~\bibnamefont {Aspuru-Guzik}},\ }\bibfield  {title} {\bibinfo {title} {{Expressibility and Entangling Capability of Parameterized Quantum Circuits for Hybrid Quantum-Classical Algorithms}},\ }\href {https://doi.org/10.1002/qute.201900070} {\bibfield  {journal} {\bibinfo  {journal} {Adv. Quantum Technol.}\ }\textbf {\bibinfo {volume} {2}},\ \bibinfo {pages} {1900070} (\bibinfo {year} {2019})}\BibitemShut {NoStop}%
\end{thebibliography}%

\onecolumngrid
\appendix
\setcounter{equation}{0}
\setcounter{table}{0}
\makeatletter

\renewcommand{\theequation}{A\arabic{equation}}
\renewcommand{\thefigure}{\arabic{figure}}

%%%%%%%%%%%%%%%%%%%%%%%%%%%%%%%%%%%%%%%%%%%%%%%%
%%%%%%%%%%%%%%%%%%%%%%%%%%%%%%%%%%%%%%%%%%%%%%%%
%%%%%%%%%%%%%%%%%%%%%%%%%%%%%%%%%%%%%%%%%%%%%%%%
%\section*{Appendix}

\section{Proof of Theorem~\ref{thm:tight_upper_bound_gen_error} (Upper bound for the general noise model)}\label{app:tighter_UB_proof}
\textbf{Theorem~\ref{thm:tight_upper_bound_gen_error} statement.}
For probabilistic errors $\mE$ of the form:
$\mE(\rho) = p \mN(\rho) + (1-p) \rho$, where the channel $\mN$ is such that $\|\mN(X)\|_2\leq \|X\|_2 \;\forall\, X\in \ml(\mathcal{H})$ where $\ml$ denotes the set of linear operators in a Hilbert space $\mathcal{H}$, we show that the upper bound in the error $\epsilon := \|\Delta\dot{\theta}\|$ is given by

    \eqs{
\frac{\epsilon}{\|\dot{\theta} \|} \leq {\rm cond}(M)
\left(\left(1- (1-p)^{2N}\right)\left( 1+\frac{N}{2\|M\|} \right) + (1- (1-p)^N)\left( 1 + \frac{\sqrt{N}}{\sqrt{2\|Y\|}}\right)\right)
}

for the probability of noise $p$ in the following range
\eqs{0\leq p\leq 1- \left( 1- \frac{2N^2\delta}{N+2\|M\|}\right)^{\frac{1}{2N}}.}

\begin{proof}
We know that 
\eqs{
\begin{split}
    M_{k,q}^{\rm err} &= \sum_{i,j}  r^*_{k,i}r_{q,j}\tr\left[\left(\mR_{k,i}\rho_0\right)^{\dagger}\left(\mR_{q,j}\rho_0 \right)\right],\; {\rm and}\\
    Y_k^{\rm err} &= - \sum_{i}\{r_{k,i} \tr\[(\mR_{k,i}\rho_0)(\Tilde{\rho} H + H\Tilde{\rho}) \]\}
\end{split}}
 where $\Tilde{\rho}$ denotes the erroneous $\rho$ that we get from the circuit subjected to errors.
%In the above equation, we can denote $\rho V_j$ as $\bar{\mR}_{N+1, j}\rho_0 = S_j(R_{N}\cdots R_1(\rho_0))T_j^{\dagger}$
%where $S_j = I$ and $T_j^{\dagger} = V_j$,
%and $\mR_{k,i}\rho_0 = \mR_{N+1}\mR_N\cdots \mR_{k+1}[S_{k,i}()$
%Then, we have
%\eqs{\tr[(\mR_{k,i}\rho_0)(\mR_{N,j}\rho_0)]}
Note that when $\theta$ is real, then $M_{k,q}$ and $Y_k$ are also real.

For single-qubit gates, we have the following forms of $r$ coefficients and the $S$ and $T$ matrices. Suppose we are differentiating at the $k$-th gate with respect to the $k$-th parameter, then
\eqs{
r_{k,1} = -\frac{i}{2}\,,&\;S_{k,1} = P_k R_k\,,\; T_{k,1} = R_k\\
r_{k,2} = \frac{i}{2}\,,&\;S_{k,2} = R_k\,,\; T_{k,2} = P_k R_k
}
Note, in a multi-qubit circuit, $P_k$ is a compact notation of the form $I\otimes\cdots\otimes I\otimes P\otimes I\otimes \cdots \otimes I$ where $P$ is the Pauli matrix that occurs in the $k$-th rotation gate.
If there is a two-qubit gate at the $k$-th position, then the $r$ coefficients and the $S$ and $T$ matrices would have the following form
\eqs{
\begin{split}
r_{k,1} = \frac{i}{2}\,,&\;S_{k,1} = |0\rangle\langle 0|\otimes I\,,\; T_{k,1} = |1\rangle\langle 1| \otimes P_kR_k  \\
r_{k,2} = -\frac{i}{2}\,,&\;S_{k,2} = |1\rangle\langle 1| \otimes P_kR_k\,,\; T_{k,2} = |0\rangle\langle 0| \otimes I  \\
r_{k,3} = -\frac{i}{2}\,,&\;S_{k,3} = |1\rangle\langle 1| \otimes P_kR_k\,,\; 
T_{k,3} = |1\rangle\langle 1| \otimes R_k  \\
r_{k,4} = \frac{i}{2}\,,&\;S_{k,4} = |1\rangle\langle 1| \otimes R_k\,,\; T_{k,4} = |1\rangle\langle 1| \otimes R_kP_k
\end{split}
}
or equivalently,
\eqs{
\begin{split}
r_{k,1} = \frac{i}{2}\,,\;&\;S_{k,1} = |0\rangle\langle 0|\otimes I + |1\rangle\langle 1| \otimes R_k\,,%\\
%&
\; T_{k,1} = |1\rangle\langle 1| \otimes P_kR_k\\
r_{k,2} = -\frac{i}{2}\,,\;&\;S_{k,2} = |1\rangle\langle 1| \otimes P_kR_k\,,\; T_{k,2} = |0\rangle\langle 0| \otimes I + |1\rangle\langle 1|\otimes R_k
\end{split}
}

When the gates at the $k$-th and the $q$-th are both single-qubit or two-qubit, then,
\eqs{\label{eq:r_values}
\begin{split}
    r^*_{k,1}r_{q,1} &= r^*_{k,2}r_{q,2} = 1/4\,\; \textrm{\, and,}\\
    r^*_{k,2}r_{q,1} &= r^*_{k,1}r_{q,2} = - 1/4\,.
\end{split}
}
If one of the two gates is a single-qubit gate and the other one is a two qubit gate, then the signs gets reversed in the above equation.

First let us only consider single-qubit gates at the $k$-th and $q$-th sites in the circuit.
Then,
\eqs{\label{eqs:T_kiqj_dephasing}
\begin{split}
    \tr[(\mR_{k,i}\rho_0)^{\dagger}(\mR_{q,j}\rho_0)] = \tr\Big[&\Big( p^{N-k+1}\mD\circ \mR_N \circ\mD\circ\cdots\mD\mR_{k+1}\mD(T_{k,i}\Tilde{\rho}_{k-1}S^{\dagger}_{k,i}) \\
    &\; +p^{N-k}(1-p)({\rm all\;} N-k {\rm \;combinations\;of\;}\mD{\rm \;after \;gates})(T_{k,i}\Tilde{\rho}_{k-1}S^{\dagger}_{k,i})\\
    &\;+ p^{N-k-1}(1-p)^2({\rm all\;} N-k-1 {\rm \;combinations\;of\;}\mD{\rm \;after \;gates})(T_{k,i}\Tilde{\rho}_{k-1}S^{\dagger}_{k,i})\\
    &\;+ \cdots + (1-p)^{N-k+1}(\mR_N\cdots\mR_{k+1}(T_{k,i}\Tilde{\rho}_{k-1}S^{\dagger}_{k,i}))\Big)\\
    &\Big( p^{N-q+1}\mD\circ \mR_N \circ\mD\circ\cdots\mD\mR_{q+1}\mD(S_{q,j}\Tilde{\rho}_{q-1}T^{\dagger}_{q,j}) \\
    &\; +p^{N-q}(1-p)({\rm all\;} N-q {\rm \;combinations\;of\;}\mD{\rm \;after \;gates})(S_{q,j}\Tilde{\rho}_{q-1}T^{\dagger}_{q,j})\\
    &\;+ p^{N-q-1}(1-p)^2({\rm all\;} N-q-1 {\rm \;combinations\;of\;}\mD{\rm \;after \;gates})(S_{q,j}\Tilde{\rho}_{k-1}T^{\dagger}_{q,j})\\
    &\;+ \cdots + (1-p)^{N-q+1}(\mR_N\cdots\mR_{q+1}(S_{q,j}\Tilde{\rho}_{q-1}T^{\dagger}_{q,j}))\Big)
    \Big]
\end{split}
}
where $\Tilde{\rho}_{k-1}$ has the following form (expression of $\Tilde{\rho}_{q-1}$ will follow similarly) 
\eqs{
\begin{split}
    \Tilde{\rho}_{k-1} &= p^{k-1}\mD\circ\mR_{k-1}\circ \mD\circ\mR_{k-2}\circ\mD\circ\cdots\circ\mR_2\circ\mD\mR_1(\rho_0)\\
    &\;+p^{k-2}(1-p)(({\rm sum\;of\;all\;possible\;}k-2{\rm \;combinations\;of\;}\mD{\rm\;after\;gates})(\rho_0))\\
    &\;+p^{k-3}(1-p)^2(({\rm sum\;of\;all\;possible\;}k-3{\rm \;combinations\;of\;}\mD{\rm\;after\;gates})(\rho_0))\\
    &\;+\cdots+(1-p)^{k-1}\mR_{k-1}\cdots\mR_1(\rho_0)\,.
\end{split}
}
We know that
\eqs{M_{k,q}^{\rm err} = \frac{1}{4}
               \left(
               \tr[(\mR_{k,1}\rho_0)^{\dagger} (\mR_{q,1}\rho_0)]
              -\tr[(\mR_{k,1}\rho_0)^{\dagger}(\mR_{q,2}\rho_0)]
              -\tr[(\mR_{k,2}\rho_0)^{\dagger}(\mR_{q,1}\rho_0)]
              +\tr[(\mR_{k,2}\rho_0)^{\dagger}(\mR_{q,2}\rho_0)]\right)
}
which can be expressed as
\eqs{\label{eqs:general_Mkq_dephasing}
\begin{split}
M_{k,q}^{\rm err} = \frac{1}{4}
               \Big(& \sum_l p_l\tr[\mN_l(\Tilde{\rho}_{_{k,l}}P_k - P_k\Tilde{\rho}_{_{k,l}})\mM_l(P_q\Tilde{\rho}_{_{q,l}} - \Tilde{\rho}_{_{q,l}}P_q)] \\
               &\;+ (1-p)^{2N}\tr\left[\left(\mR_{N}\cdots\mR_{k+1}(\rho_kP_k -P_k\rho_k)\right)\left(\mR_{N}\cdots\mR_{q+1}(P_q\rho_q -\rho_qP_q)\right)\right]\Big)
\end{split}
}
where $p_l$ are probabilities which are of the form $p^a (1-p)^b$ such that $\sum_l p_l = 1- (1-p)^{2N}$.
The first term in the RHS of the Eq.~\eqref{eqs:general_Mkq_dephasing} is a general expression of writing the terms where $\mD$ occurs atleast once either in $\mN_l$, $\mM_l$, $\Tilde{\rho}_{_{k,l}}$, and $\Tilde{\rho}_{_{q,l}}$.
That is, $\mN_l$ is of the form $\mD^{a_N}\circ\mR_N\circ\mD^{a_{N-1}}\circ\mR_{N-1}\circ\cdots\circ\mD^{a_{k+1}}\circ\mR_{k+1}$
where $a_i={0,1}$, and $\Tilde{\rho}_{_{k,l}}$ has the form $\mD^{b_{k-1}}\circ\mR_{k-1}\circ \mD^{b_{k-1}}\circ\mR_{k-2}\circ\cdots\circ\mD^{b_1}\mR_1(\rho_0)$ where $b_i = {0,1}$ and the probabilities have been taken care of in $p_l$, and similarly for $\mM_l$ and $\Tilde{\rho}_{_{q,l}}$.
Using the Cauchy-Schwarz inequality we can write 
\eqs{\label{eq:using_Cauchy_Shwarz}
\begin{split}
\tr[\mN_l(\Tilde{\rho}_{_{k,l}}P_k &- P_k\Tilde{\rho}_{_{k,l}})\mM_l(P_q\Tilde{\rho}_{_{q,l}} - \Tilde{\rho}_{_{q,l}}P_q)]\\
&\leq
\sqrt{\tr[\mN_l(\Tilde{\rho}_{_{k,l}}P_k - P_k\Tilde{\rho}_{_{k,l}})\mN_l(P_k\Tilde{\rho}_{_{k,l}} - \Tilde{\rho}_{_{k,l}}P_k)]}
\sqrt{\tr[\mM_l(P_q\Tilde{\rho}_{_{q,l}} - \Tilde{\rho}_{_{q,l}}P_q)\mM_l(\Tilde{\rho}_{_{q,l}}P_q - P_q\Tilde{\rho}_{_{q,l}})]}\\
&=\|\mN_l(\Tilde{\rho}_{_{k,l}}P_k - P_k\Tilde{\rho}_{_{k,l}}) \|_2 \| \mM_l(P_q\Tilde{\rho}_{_{q,l}} - \Tilde{\rho}_{_{q,l}}P_q)\|_2 
\end{split}
}
The first inequality follows since in the LHS we are taking the product of two anti-Hermitian matrices and if we multiply both the matrices by $i$ and take a minus sign out of trace, we essentially get LHS to be equal to be less than or equal to the trace of two Hermitian matrices and then when we apply the Cauchy-Schwarz inequality, we get the above bound.
{Furthermore, it also holds that
\eqs{\label{eq:using_Cauchy_Shwarz_again}
\tr[\mN_l(\Tilde{\rho}_{_{k,l}}P_k - P_k\Tilde{\rho}_{_{k,l}})\mM_l(\Tilde{\rho}_{_{q,l}}P_q-P_q\Tilde{\rho}_{_{q,l}} )]
\leq
\|\mN_l(\Tilde{\rho}_{_{k,l}}P_k - P_k\Tilde{\rho}_{_{k,l}}) \|_2 \| \mM_l(P_q\Tilde{\rho}_{_{q,l}} - \Tilde{\rho}_{_{q,l}}P_q)\|_2 
}
and therefore,
\eqs{|\tr[\mN_l(\Tilde{\rho}_{_{k,l}}P_k - P_k\Tilde{\rho}_{_{k,l}})\mM_l(P_q\Tilde{\rho}_{_{q,l}} - \Tilde{\rho}_{_{q,l}}P_q)]|\leq
\|\mN_l(\Tilde{\rho}_{_{k,l}}P_k - P_k\Tilde{\rho}_{_{k,l}}) \|_2 \| \mM_l(P_q\Tilde{\rho}_{_{q,l}} - \Tilde{\rho}_{_{q,l}}P_q)\|_2 }
}
Next, note that for any operator, say $X$,
we have
\eqs{\label{eq:inequality_by_requirement}
\|\mN_l(X)\|_2\leq \|X\|_2 = \|\mR_N\circ\mR_{N-1}\circ\cdots\circ\mR_{k+1}(X) \|_2}
Now, let us denote $\mR_N\circ\mR_{N-1}\circ\cdots\circ\mR_{k+1}$ as $\mR$, then have the following
\eqs{
\|\mR(X)\|_2 &= \|\mR(\Tilde{\rho}_{_{k,l}}P_k - P_k\Tilde{\rho}_{_{k,l}})\|_2\\
&= \|\Tilde{\rho}_{_{k,l}}P_k - P_k\Tilde{\rho}_{_{k,l}}\|_2\\
&=\|P_k\Tilde{\rho}_{_{k,l}}P_k - \Tilde{\rho}_{_{k,l}}\|_2\\
&\leq \sqrt{2}\label{eq:bound_on_rhoP-prho}
}

\begin{comment}
Since $P_k$ is an $n$-qubit Pauli operator of the form $I\otimes I\otimes\cdots\otimes\sigma_k\otimes\cdots \otimes I$
where $\sigma_k$ denotes the Pauli associated with the $k$-th rotation gate, the $\ell-2$ norm of $P_k$ is equal to $\sqrt{d}$ where $d=2^n$.

Now, if we suppose
$\Tilde{\rho}_{k,l} = \mU(\mD(\Tilde{\rho}_{x-1,l})$
where $\mU(\cdot)=U(\cdot)U^{\dagger}$ represents the unitary $\mR_{k-1}\cdots\mR_x(\cdot)$, then
\eqs{
\|P_k\Tilde{\rho}_{_{k,l}}P_k - \Tilde{\rho}_{_{k,l}}\|_2 &= \|P_k\mU(\mD(\Tilde{\rho}_{_{x-1,l}}))P_k - \mU(\mD(\Tilde{\rho}_{_{x-1,l}}))\|_2\\
&=\left\|V\left(\mD(\Tilde{\rho}_{_{x-1,l}})\right)V^{\dagger}-\mD(\Tilde{\rho}_{_{x-1,l}})\right\|_2
}
where $V = U^{\dagger}P_k U$.
Since 
$$\mD(A) = \frac{1}{2}(A + ZAZ) $$
we can write
\eqs{\left\|V\left(\mD(\Tilde{\rho}_{_{x-1,l}})\right)V^{\dagger}-\mD(\Tilde{\rho}_{_{x-1,l}})\right\|_2 &= 
\frac{1}{2}\|V(\Tilde{\rho}_{_{x-1,l}} + Z\Tilde{\rho}_{_{x-1,l}}Z)V^{\dagger} - (\Tilde{\rho}_{_{x-1,l}} + Z\Tilde{\rho}_{_{x-1,l}}Z) \|_2\\
&\leq \frac{1}{2}\|V\Tilde{\rho}_{_{x-1,l}}V^{\dagger}-\Tilde{\rho}_{_{x-1,l}}\|_2 + \frac{1}{2}\|VZ\Tilde{\rho}_{_{x-1,l}}ZV^{\dagger}-Z\Tilde{\rho}_{_{x-1,l}}Z\|_2\\
&= \frac{1}{2}\|V\Tilde{\rho}_{_{x-1,l}}V^{\dagger}-\Tilde{\rho}_{_{x-1,l}}\|_2 + \frac{1}{2}\|ZVZ\Tilde{\rho}_{_{x-1,l}}ZV^{\dagger}Z-\Tilde{\rho}_{_{x-1,l}}\|_2
}
\end{comment}

Therefore, for single-qubit gates (at $k$-th and $q$-th sites) we have 
\eqs{|\tr[\mN_l(\Tilde{\rho}_{_{k,l}}P_k - P_k\Tilde{\rho}_{_{k,l}})\mM_l(P_q\Tilde{\rho}_{_{q,l}} - \Tilde{\rho}_{_{q,l}}P_q)]|\leq 2\,.}

Now, let us consider that the gates at the $k$-th site and $q$-site are two-qubit gates.
Let $W_k = |0\rangle\langle0 | \otimes I + |1\rangle\langle1| \otimes R_k$ 
and $W_q = |0\rangle\langle0 | \otimes I + |1\rangle\langle1| \otimes R_q$ be two unitary matrices.
Then $M^{\rm err}_{k,q}$ can be expressed as follows
\eqs{
\label{eq:single_qubit_general_Mkq_dephasing}
\begin{split}
M^{\rm err}_{k,q} &= \frac{1}{4}\left(\tr[(\mR_{k,1}\rho_0-\mR_{k,2}\rho_0)^{\dagger}(\mR_{q,1}\rho_0-\mR_{q,2}\rho_0)]\right)\\
&=\frac{1}{4}\Big(\sum_l p_l \tr\left[\mN_l\left(T_{k,1}\Tilde{\rho}_{k-1}S^{\dagger}_{k,1} - T_{k,2}\Tilde{\rho}_{k-1}S^{\dagger}_{k,2}\right)
\mM_l\left(S_{q,1}\Tilde{\rho}_{q-1}T^{\dagger}_{q,1} - S_{q,2}\Tilde{\rho}_{q-1}T^{\dagger}_{q,2}\right)\right] +\\
&(1-p)^{2N}\tr\left[\left(\mR_{N}\cdots\mR_{k+1}\left(T_{k,1}{\rho}_{k-1}S^{\dagger}_{k,1} - T_{k,2}{\rho}_{k-1}S^{\dagger}_{k,2}\right)\right)\left(\mR_{N}\cdots\mR_{q+1}\left(S_{q,1}{\rho}_{q-1}T^{\dagger}_{q,1} - S_{q,2}{\rho}_{q-1}T^{\dagger}_{q,2}\right)\right)\right]
\Big)\\
&=\frac{1}{4}\left(\sum_l p_l \tr\left[\mN_l\left(T_{k,1}\Tilde{\rho}_{k-1}S^{\dagger}_{k,1} - T_{k,2}\Tilde{\rho}_{k-1}S^{\dagger}_{k,2}\right)
\mM_l\left(S_{q,1}\Tilde{\rho}_{q-1}T^{\dagger}_{q,1} - S_{q,2}\Tilde{\rho}_{q-1}T^{\dagger}_{q,2}\right)\right]\right) + (1-p)^{2N} M_{k,q}
\end{split}
}
Also, for two-qubits gates we get
\eqs{
\begin{split}
    S_{k,1}\Tilde{\rho}_{k-1}T^{\dagger}_{k,1} - S_{k,2}\Tilde{\rho}_{k-1}T^{\dagger}_{k,2} &= (|0\rangle\langle 0|\otimes I + |1\rangle\langle 1|\otimes R_k)\Tilde{\rho}_{k-1}(|1\rangle\langle 1|\otimes R_k^{\dagger}P_k) - 
    ( |1\rangle\langle 1|\otimes P_k R_k)\Tilde{\rho}_{k-1}(|0\rangle\langle 0|\otimes I +|1\rangle\langle 1|\otimes R_k^{\dagger})\\
    &=\Tilde{\rho}_k(|1\rangle\langle 1|\otimes P_k) - (|1\rangle\langle 1|\otimes P_k) \Tilde{\rho}_k
\end{split}
}
Similarly, we can obtain the expression for the gate at $q$-th site as
\eqs{S_{q,1}\Tilde{\rho}_{q-1}T^{\dagger}_{q,1} - S_{q,2}\Tilde{\rho}_{q-1}T^{\dagger}_{q,2} = \Tilde{\rho}_q (|1\rangle\langle 1| \otimes P_q) - (|1\rangle \langle 1|\otimes P_q)\Tilde{\rho}_q}
Therefore, we can write
\eqs{\label{eqs:two_qubit_general_Mkq_dephasing}
\begin{split}
M_{k,q}^{\rm err} =\frac{1}{4}\left(\sum_l p_l \tr\big[\mN_l\left(\left(|1\rangle\langle 1|\otimes P_k\right) \Tilde{\rho}_k -\Tilde{\rho}_k\left(|1\rangle\langle 1|\otimes P_k\right) \right)
\mM_l\left(\Tilde{\rho}_q \left(|1\rangle\langle 1| \otimes P_q\right) - \left(|1\rangle \langle 1|\otimes P_q\right)\Tilde{\rho}_q\right)\big]\right) + (1-p)^{2N} M_{k,q} 
\end{split}
}
Let us now simplify the above expression by finding an upper bound to the first terms in RHS
\eqs{
\begin{split}
|\tr\big[\mN_l\big(\left(|1\rangle\langle 1|\otimes P_k\right) &\Tilde{\rho}_k -\Tilde{\rho}_k\left(|1\rangle\langle 1|\otimes P_k\right) \big)
\mM_l\big(\Tilde{\rho}_q \left(|1\rangle\langle 1| \otimes P_q\right) - \left(|1\rangle \langle 1|\otimes P_q\right)\Tilde{\rho}_q\big)\big]|\\
&\leq \|\mN_l\big(\left(|1\rangle\langle 1|\otimes P_k\right) \Tilde{\rho}_k -\Tilde{\rho}_k\left(|1\rangle\langle 1|\otimes P_k\right) \big) \|_2 
\| \mM_l\big(\Tilde{\rho}_q \left(|1\rangle\langle 1| \otimes P_q\right) - \left(|1\rangle \langle 1|\otimes P_q\right)\Tilde{\rho}_q\big)\|_2
\end{split}
}
Moreover, we have
\eqs{
\begin{split}
    \|\mN_l\big(\left(|1\rangle\langle 1|\otimes P_k\right) \Tilde{\rho}_k -\Tilde{\rho}_k\left(|1\rangle\langle 1|\otimes P_k\right) \big) \|_2 &\leq \|\left(|1\rangle\langle 1|\otimes P_k\right) \Tilde{\rho}_k -\Tilde{\rho}_k\left(|1\rangle\langle 1|\otimes P_k\right) \|_2\\
    &= \sqrt{\tr\left[\left(\left(|1\rangle\langle 1|\otimes P_k\right) \Tilde{\rho}_k -\Tilde{\rho}_k\left(|1\rangle\langle 1|\otimes P_k\right)\right)\left(\Tilde{\rho}_k\left(|1\rangle\langle 1|\otimes P_k\right) - \left(|1\rangle\langle 1|\otimes P_k\right) \Tilde{\rho}_k \right)\right]}\\
    &=\sqrt{2}\sqrt{\tr[\Tilde{\rho}^2(|1\rangle\langle 1|\otimes I)] - \tr[\Tilde{\rho}(|1\rangle\langle 1|\otimes P_k)\Tilde{\rho}(|1\rangle\langle 1|\otimes P_k)]}\\
    &\leq \sqrt{2}
\end{split}
}
Therefore,  for two-qubit gates (at $k$-th and $q$-th sites) we have 
\eqs{|\tr\big[\mN_l\big(\left(|1\rangle\langle 1|\otimes P_k\right) &\Tilde{\rho}_k -\Tilde{\rho}_k\left(|1\rangle\langle 1|\otimes P_k\right) \big)
\mM_l\big(\Tilde{\rho}_q \left(|1\rangle\langle 1| \otimes P_q\right) - \left(|1\rangle \langle 1|\otimes P_q\right)\Tilde{\rho}_q\big)\big]|\leq 2}

Thus, we can bound $M^{\rm err}_{k,q}$ as
\eqs{
{-(1- (1-p)^{2N}))\frac{1}{2} + (1-p)^{2N} M_{k,q} \leq}
M^{\rm err}_{k,q}\leq (1- (1-p)^{2N}))\frac{1}{2} + (1-p)^{2N} M_{k,q}
}
which implies
{\eqs{
|M^{\rm err}_{k,q} - (1-p)^{2N} M_{k,q}|\leq \frac{1 - (1-p)^{2N}}{2}
}
Therefore, we can say that for the matrix $M^{\rm err} - (1-p)^{2N}M$, we have
\eqs{
\|M^{\rm err} - (1-p)^{2N}M \|_2 \leq (1 - (1-p)^{2N})\frac{N}{2}
}
Using this, we can get an upper bound on $\|\Delta M\|$ as given below
\eqs{
\|M^{\rm err} - M \| &= \|M^{\rm err} -(1-p)^{2N}M + (1-p)^{2N}M - M \|\\
&\leq \|M^{\rm err} -(1-p)^{2N}M\| + (1-(1-p)^{2N})\|M \|\\
&\leq (1-(1-p)^{2N})\left(\frac{N}{2} + \|M\|\right)
}
Hence,
\eqs{\frac{\|\Delta M \|}{\|M\|}\leq \left(1- (1-p)^{2N}\right)\left( \frac{N}{2\|M\|}+1 \right)\,.\label{eq:M_true_upper_bound}
}
}
\begin{comment}
Therefore,
\eqs{\label{eq:upperbound_Merr}
M^{\rm err} \leq \frac{1- (1-p)^{2N}}{2}|\Tilde{+}\rangle\langle \Tilde{+}| + (1-p)^{2N} M
}
where $|\Tilde{+}\rangle\langle \Tilde{+}|$ denotes the $N\times N$ matrix with all entries being one.

Now, using Eqs.~\eqref{eq:using_Cauchy_Shwarz},\eqref{eq:inequality_by_requirement},\eqref{eq:bound_on_rhoP-prho},\eqref{eq:single_qubit_general_Mkq_dephasing},\eqref{eqs:two_qubit_general_Mkq_dephasing}, we get that $M_{k,q}\leq 1/2$. %and so, we can write
%\eqs{M\leq \frac{1}{2}|\Tilde{+}\rangle\langle\Tilde{+}|\,.}
%This also implies that $\|M\|\leq N/2$.
Now, using the above inequality, we have the upper bound on $\Delta M$ to be
\eqs{
\Delta M:= M^{\rm err} - M &\leq \frac{1- (1-p)^{2N}}{2}|\Tilde{+}\rangle\langle \Tilde{+}| + (1-p)^{2N} M - M\\
&= (1-(1-p)^{2N})(\frac{1}{2}|\Tilde{+}\rangle\langle \Tilde{+}| - M) %\\
%&\leq \frac{1-(1-p)^{2N}}{2}|\Tilde{+}\rangle\langle \Tilde{+}|
}

So, we have 
\eqs{\label{eq:deltaM/M_dephasing}
\frac{\|\Delta M\|}{\|M\|}\leq \left(1- (1-p)^{2N}\right) \frac{\|\frac{1}{2}|\Tilde{+}\rangle\langle \Tilde{+}| - M \|}{\|M\|}
}  
\end{comment}

Similarly, we can find the upper bound on $Y^{\rm err}_{k,q}$ to be
\eqs{
Y^{\rm err}_{k,q}\leq (1- (1-p)^{N}))\frac{1}{\sqrt{2}} + (1-p)^{N} Y_{k,q}
}
{
and from this we get
\eqs{
|Y^{\rm err}_{k,q} - (1-p)^N Y_{k,q}| \leq (1- (1-p)^N)\frac{1}{\sqrt{2}}
}
Following steps similar to the upper bound in Eq.~\eqref{eq:M_true_upper_bound}, we get the upper bound on $\|\Delta Y\|$ to be
\eqs{
\frac{\|\Delta Y \|}{\|Y\|}\leq (1- (1-p)^N)\left( 1 + \frac{\sqrt{N}}{\sqrt{2\|Y\|}}\right)\label{eq:true_Y_upperbound}
}
}
\begin{comment}
and the vector $Y$ can be bounded as 
\eqs{
Y^{\rm err} \leq \frac{1- (1-p)^{N}}{\sqrt{2}}|\Tilde{+}\rangle + (1-p)^{N} Y.
}
and we will have the following inequality
\eqs{\label{eq:deltaY/Y_dephasing}
\frac{\|\Delta Y\|}{\|Y\|}\leq \left(1- (1-p)^{N}\right)\frac{\|\frac{1}{\sqrt{2}}|\Tilde{+}\rangle - Y\|}{\|Y\|}\,.
}
    
\end{comment}
% and that $\|Y\|_2 \leq \| \frac{1}{\sqrt{2}}|\Tilde{+}\rangle\|_2$ can be derived in the same way as that of $M$.

Finally, to derive upper bounds on the error in $\dot{\theta}$ we utilize the formalism of studying perturbations to matrix linear equations to provide bounds on the solution of our matrix equation $M \dot{\theta} = Y$ where both $M$ and $Y$ are perturbed. The ($p$-)condition number (denoted ${\rm cond}_p$) of a matrix $Z$ is a defined as
\eqs{{\rm cond}_p(Z):= \|Z\|_p \|Z^{-1}\|_p}
where $\|\cdot\|_p$ represents the matrix $p$-norm. 
If $Z$ is not invertible, then ${\rm cond}_p(Z) = \infty$.
In our work, we use $\|\cdot\|$ to mean $\|\cdot\|_2$ defined as $\|Z\|_2 := \sqrt{\tr(Z^*Z)}$ (also known as the Frobenius norm), and ${\rm cond}(Z)$ to mean ${\rm cond}_2(Z)$, unless otherwise stated. If we consider a perturbation in both $M$ and $Y$ of the form
\eqs{
(M+\Delta M)(\Delta \dot{\theta} + \dot{\theta})= Y+\Delta Y
}
then the solution to the perturbed system obeys the following inequality \cite{Higham2002}
\eqs{
\label{eqn:CondNumbIneq}
\frac{\|\Delta \dot{\theta}\|}{\|\dot{\theta}\|} \leq \frac{{\rm cond}(M)}{1-{\rm cond}(M)\frac{\|\Delta M\|}{\|M\|}} \left ( \frac{\|\Delta M\|}{\|M\|}+\frac{\|\Delta Y\|}{\|Y\|} \right )\,.
}
{Using the above inequality, we calculate the upper bounds under general noise model and show the impact of generalized unitary errors in any given circuit amounts to perturbing $M$ and $Y$.}

By plugging in Eqs.~\eqref{eq:M_true_upper_bound} and~\eqref{eq:true_Y_upperbound} into equation \eqref{eqn:CondNumbIneq}, we can bound the error, $\epsilon := \|\Delta\dot{\theta}\|$ as
\eqs{\label{eq:initial_UB}
\epsilon \leq \frac{\|\dot{\theta} \|{\rm cond}(M)}{1-{\rm cond}(M)\left(\left(1- (1-p)^{2N}\right)\left( 1+\frac{N}{2\|M\|} \right)\right)}
\left(\left(1- (1-p)^{2N}\right)\left( 1+\frac{N}{2\|M\|} \right) + (1- (1-p)^N)\left( 1 + \frac{\sqrt{N}}{\sqrt{2\|Y\|}}\right)\right)
}
{which is valid under the constraint that}
\eqs{\label{eq:constraint_on_cond_generic}
1\leq {\rm cond}(M)\leq \frac{2\|M\|}{(1 - (1-p)^{2N})(N+2\|M\|)},}
{for a given probability of noise where cond$(\cdot)$ represents the condition number of the matrix and $N$ is the total number of parameters in the circuit.
Equivalently, the above constraint can be expressed as a constraint on probability provided the condition number of M is known as follows}
\eqs{\label{eq:constraint_on_prob_generic}
0\leq p \leq 1- \left( 1 - \frac{2\|M\|}{{\rm cond}(M)(N+2\|M\|)}\right)^{\frac{1}{2N}},}
{The above constraints arise due to the fact that the right hand side of Eq.~\eqref{eq:initial_UB} needs to be positive}.
{The constraint on condition number can be understood as follows: for a given probability of noise, the upper bound on the error can be found for condition numbers between one and the bound in} Eq.~\eqref{eq:constraint_on_cond_generic}. Similarly, we can interpret the constraint on the allowed probability range. 
Further analyzing the constraint on the condition number implies that as the probability of noise decreases, the upper bound can be found for matrices with larger condition numbers, and therefore, as the noise probability tends to zero, the upper bound can be found for almost all the cases (as the bound on the condition number tends to infinity).
{Also note that the blow-out in the error as probability of noise is increasing is coming from the fundamental result of perturbation theory and not from noise in QITE algorithm itself. Therefore, it is crucial to find tighter bounds that are valid for larger probabilities of noise.}

{To find tighter bounds, we need to impose certain physical constraints or assumptions. The above general result is derived without any constraint.
Now, if we assume that the contribution of $\Delta M \Delta \dot{\theta}$ to $\Delta Y$ in the perturbed equation $(M + \Delta M)(\dot{\theta} + \Delta \dot{\theta}) = Y + \Delta Y$ can be considered negligible when $\| \Delta M\|\leq N^2 \delta$ and $\|\Delta \dot{\theta}\|\leq N \delta$ for some fixed very small $\delta << 1$, then we can find bounds tighter than the one presented above.}

{So, for the above constraint to work, we need the following to hold:
\eqs{\label{eq:assumption_on_DeltaM}
\|\Delta M\|\leq \|M\|(1- (1-p)^{2N})\left( 1 + \frac{N}{2\|M\|}\right)\leq N^2 \delta
}
where the first inequality is coming form Eq.~\eqref{eq:M_true_upper_bound} and the second inequality is imposed so that the contribution of $\Delta M \Delta \dot{\theta}$ can be considered negligible.
By this assumption we then have
\eqs{
M\Delta\dot{\theta} + \Delta M\dot{\theta} = \Delta Y
}
from which we get
\eqs{
\|\Delta\dot{\theta}\|\leq \|M^{-1} \|(\|\Delta Y\| + \|M\|\|\dot{\theta}\|)\,.
}
From the above equation and by using $\|Y\|\leq \|M\|\|\dot{\theta}\|$, we get
\eqs{
\frac{\|\Delta\dot{\theta}\|}{\|\dot{\theta}\|}&\leq {\rm cond}(M)\left(\frac{\|\Delta M\|}{\|M\|} + \frac{\|\Delta Y\|}{\| Y\|}\right)\\
&\leq {\rm cond}(M)
\left(\left(1- (1-p)^{2N}\right)\left( 1+\frac{N}{2\|M\|} \right) + (1- (1-p)^N)\left( 1 + \frac{\sqrt{N}}{\sqrt{2\|Y\|}}\right)\right)
}
which is valid when the probability of noise is in the following range:
\eqs{0\leq p\leq 1- \left( 1- \frac{2N^2\delta}{N+2\|M\|}\right)^{\frac{1}{2N}}.}
The above constraint on M is straightforward to get from Eq.~\eqref{eq:assumption_on_DeltaM}.
}
\end{proof}

%\renewcommand{\theequation}{B\arabic{equation}}
%\section{Upper Bound in error for dephasing noise}\label{app:dephasing_noise}
%gg
\renewcommand{\theequation}{B\arabic{equation}}
\section{Proof of Theorem~\ref{thm:upper_bound_depolarizing} (Error under depolarizing noise)}
\label{app:C}

\textbf{Theorem~\ref{thm:upper_bound_depolarizing} statement.}
For partially depolarizing channels, the upper bound in the error $\epsilon := \|\Delta\dot{\theta}\|$ is given by
\eqs{
\frac{\epsilon}{\|\dot{\theta}\|}  = \frac{1-(1-p)^N}{(1-p)^N}.
}
\begin{proof}
The action of the depolarizing channel is given by
\eqs{\mE(X) = p\tr[X]\frac{I}{d} + (1-p)X\,.}

Assuming that the depolarizing noise affects the circuit after the application of each gate, we get the following expression for $\tr\left[\left(\mR_{k,i}\rho_0\right)^{\dagger}\left(\mR_{q,j}\rho_0 \right)\right]$
\eqs{\label{eq:tr_Rki_Rqj}
\begin{split}
\tr\left[\left(\mR_{k,i}\rho_0\right)^{\dagger}\left(\mR_{q,j}\rho_0 \right)\right]
=\;& \(p(1-p)^{k-1}(a_0 + a_1p +\ldots + a_{N-k}p^{N-k})\tr[P_k\rho_{k-1}] \)\Bigg(
                            \frac{1}{d}p(1-p)^{q-1}(b_0 + b_1p+\ldots+b_{N-q}p^{N-q})\times\\
                            &\qquad\qquad\quad \tr\left[P_q\rho_{q-1}\right]
                            \Bigg) + \frac{(1-p)^N}{d} \\
+&\(1-p)^{N-k+1}(1 - (1-p)^{k-1}\)\Bigg(
                            \frac{(1-p)^{N-q+1}(1 - (1-p)^{q-1})}{d^2}\tr[\mR_N\ldots\mR_{k+1}(P_k)\mR_N\ldots\mR_{q+1}(P_q)]\\
                            &\qquad\qquad\quad+ \frac{(1-p)^N}{d}\tr\left[\mR_N\circ\cdots\circ\mR_{k+1}\left(P_k\right)\mR_N\circ\cdots\circ \mR_{q+1}\left(T_{q,j}\rho_{q-1} S^{\dagger}_{q,j}\right)\right]
                            \Bigg)\\
+&\(1-p\)^N\Bigg(
                            \frac{p}{d}(1-p)^{q-1}(b_0+b_1p+\ldots+b_{N-q}p^{N-q})\tr[P_q\rho_{q-1}]\tr[P_k\rho_{k-1}]\\
                            &\qquad\qquad\quad+ \frac{(1-p)^{N-q+1}(1-(1-p)^{q-1})}{d}\tr\left[R_N\circ\cdots\circ R_{k+1}\left(S_{k,i}\rho_{k-1}T^{\dagger}_{k,i}\right)R_N\circ\cdots\circ R_{q+1}\left(P_q\right)\right]\\
                            &\qquad\qquad\quad+ (1-p)^N\tr\left[R_N\circ\cdots\circ R_{k+1}\left(S_{k,i}\rho_{k-1}T^{\dagger}_{k,i}\right)R_N\circ\cdots\circ R_{q+1}\left(T_{q,j}\rho_{q-1} S^{\dagger}_{q,j}\right)\right]
                            \Bigg)\;.
\end{split}
}
%\end{widetext}
where $P_k(P_q)$ denotes the Pauli gate at the $k$-th($q$-th) rotation gate.
In the above expression, when we consider $p=0$, we get the ideal (non-erroneous) scenario.
We also noticed that the coefficients $a_0, a_1,\ldots,a_{N-k}$ and $b_0,b_1,\ldots,b_{N-q}$ can be found using Pascal's triangle.
To get these terms, let $n$ be the highest power in the series.
Then, the coefficients come from the $(n+2)$th row of the Pascal's triangle by ignoring the first $1$ of that row, and by alternating between the $+$ and $-$ signs.
Suppose, $n = 2$, then we have the series as $a_0 + a_1p + a_2p^2$ and from Pascal's triangle we get that $a_0 = 3$, $a_1=-3$, and $a_2 =1$.

Similarly to find $Y^{\rm err}$, when we include the depolarizing error after every gate, we get
%\begin{widetext}
\eqs{\label{eq:tr_Rki_rhoY}
\begin{split}
\tr[(\mR_{k,i}\rho_0)(\rho H + H\rho)]
&=\; \(p(1-p)^{k-1}(a_0 + a_1p +\ldots + a_{N-k}p^{N-k})\tr[P_k\rho_{k-1}] \)\left(\frac{1}{d}\tr[\rho H + H\rho]\right)\\
&\quad+\;\(1-p)^{N-k+1}(1 - (1-p)^{k-1}\)\frac{1}{d}\tr[\mR_N\cdots\mR_{k+1}(P_k)(\rho H + H\rho)]\\
&\quad+\; (1-p)^N \tr[\mR_N\cdots \mR_{k+1}(S_{k,i}\rho_{k-1}T^{\dagger}_{k,i})(\rho H + H\rho)]
\end{split}}

%\end{widetext}
Using the equations~\eqref{eq:tr_Rki_Rqj} and~\eqref{eq:tr_Rki_rhoY}, we can find the elements of $M^{\rm err}$ and $Y^{\rm err}$.
For instance, the diagonal elements of $M^{\rm err}$ (when $k$-th and $q$-th gates are single-qubit gates) can be expressed as
\eqs{M^{\rm err}_{k,k} = (1-p)^{2N}\left(\frac{1}{2} - \frac{1}{2}\tr[P_k\rho_{k-1}]^2 \right)\,.
}
The off-diagonal elements of $M^{\rm err}$, $M_{k,q}^{\rm err}$, where $k<q$, when $k$-th and $q$-th gates are single-qubit gates, can be expressed as
%%%%%%%%%%%%%%%%%%%%%%%%%%%
%Using this, we get the off-diagonal terms in $M^{\rm err}$ as
%\begin{widetext}
\eqs{
\begin{split}
M^{err}_{k,q} = \frac{(1-p)^{2n}}{4}
\Bigg(&\tr\left[R_q\circ\cdots\circ R_{k+1}\left(S_{k,1}\rho_{k-1}T^{\dagger}_{k,1}\right)\left(T_{q,1}\rho_{q-1} S^{\dagger}_{q,1}\right)\right]\\
-& \tr\left[R_q\circ\cdots\circ R_{k+1}\left(S_{k,1}\rho_{k-1}T^{\dagger}_{k,1}\right)\left(T_{q,2}\rho_{q-1} S^{\dagger}_{q,2}\right)\right]\\
-& \tr\left[R_q\circ\cdots\circ R_{k+1}\left(S_{k,2}\rho_{k-1}T^{\dagger}_{k,2}\right)\left(T_{q,1}\rho_{q-1} S^{\dagger}_{q,1}\right)\right]\\
+& \tr\left[R_q\circ\cdots\circ R_{k+1}\left(S_{k,2}\rho_{k-1}T^{\dagger}_{k,2}\right)\left(T_{q,2}\rho_{q-1} S^{\dagger}_{q,2}\right)\right]\Bigg)
\end{split}
}
%\end{widetext}
which can be written as

%\begin{widetext}
\eqs{
\begin{split}
M^{err}_{k,q} = \frac{(1-p)^{2N}}{4}
\Bigg(&\tr\left[R_q\circ\cdots\circ R_{k+1}\left(P_k\rho_{k}\right)\left(\rho_{q} P_q\right)\right]\\
-& \tr\left[R_q\circ\cdots\circ R_{k+1}\left(P_k\rho_{k}\right)\left(P_q\rho_{q} \right)\right]\\
-& \tr\left[R_q\circ\cdots\circ R_{k+1}\left(\rho_{k}P_k\right)\left(\rho_{q}P_q\right)\right]\\
+& \tr\left[R_q\circ\cdots\circ R_{k+1}\left(\rho_{k}P_k\right)\left(P_q\rho_{q} \right)\right]\Bigg)
\end{split}
}
%\end{widetext}

Since we are considering that $\rho_0$ to be a pure state, the above expression can be simplified as follows

\eqs{
M^{err}_{k,q} = \frac{(1-p)^{2N}}{4}
\Bigg(&\tr\left[R_q\circ\cdots\circ R_{k+1}(P_k)\left(\rho_{q}P_q - 2\rho_q P_q \rho_q + P_q\rho_q\right)\right]\Bigg)
}

So, the diagonal and off-diagonal elements can be expressed as
\eqs{
\begin{split}
M^{\rm err}_{k,k} &= \frac{(1-p)^{2N}}{2}\left(1 - \tr[P_k \rho_{k-1} P_k \rho_{k-1}]\right)\\
M^{err}_{k,q} &= \frac{(1-p)^{2N}}{4}
\Bigg(\tr\left[R_q\circ\cdots\circ R_{k+1}(P_k)\left(\rho_{q}P_q - 2\rho_q P_q \rho_q + P_q\rho_q\right)\right]\Bigg)
\end{split}
}
%%%%%%%%%%%%%%%%%%%%%%%%%%%%5

Since the matrix $M^{\rm err}$ is symmetric, we can find the elements $M_{k,q}^{\rm err}$ when $k>q$.
We found that even when the $k$-th and/or $q$-th gates are two-qubit gates, the same equality holds, i.e.,
\eqs{M^{\rm err}_{k,q} = (1-p)^{2N}M_{k,q}\,\label{Merr_thm2}}
which follows owing to the particular form of the depolarizing channels and due to the alternating signs in the definition of $M$ and $Y$.

Using similar calculations, the elements of $Y^{\rm err}$ can be expressed as
%\begin{widetext}
\eqs{Y_k^{\rm err} = (1-p)^N
\left(\frac{1}{2}\tr[(\mR_N\cdots\mR_{k+1}(\rho_kP_k - P_k\rho_k))\{\rho, H\}]\right)\,.
}
%\end{widetext}

%\red{The details of the controlled-rotation gates to be filled in later here}
So, we found that under partially depolarizing noise occurring after every gate, the erroneous $M$ and $Y$ scale by a factor of $(1-p)^{2N}$ and $(1-p)^N$ respectively, i.e.,
\eqs{
M^{err} &= (1-p)^{2N} M\label{eq:M_err_depol}\\
Y^{err} &= (1-p)^{N} Y \label{eq:Y_err_depol}
}

{With these equalities, given the perturbed equation
\eqs{
M^{\rm err}\dot{\theta}^{\rm err} = Y^{\rm err}
}
we can reduce it to
\eqs{(1-p)^{N}M\dot{\theta}^{\rm err} = Y\,.}
Therefore, we can write
\eqs{
(1-p)^N \dot{\theta}^{\rm err} = \dot{\theta}\,.
}
From the above expression, we get the relative error to be
\eqs{
\frac{\|\Delta \dot{\theta}\|}{\|\dot{\theta}\|} = \frac{1-(1-p)^N}{(1-p)^N}
}
}

\end{proof}

\renewcommand{\theequation}{C\arabic{equation}}
\section{Numerical Analyses}\label{app:numerics}
We performed the numerical simulation to plot the upper bound of the relative error (in the rate of change of the parameters) as a function of the probability of noise.
We plotted this error vs probability graph for the Hamiltonian of the Hydrogen molecule given in the main text.
For simulating this Hamiltonian using a general ansatz, we considered general probabilistic noise and the depolarizing noise as the noise models.
\begin{comment}
For the general probabilistic noise, we used an ansatz with five parameters (see Fig.~\ref{fig:simple_ansatz}).
We then randomly generated four sets of angles.
For these sets of angles we found the $M$ and the $Y$ matrices and the condition numbers of $M$ came out to be $231.776$, $191.5535$, $178.7267$, and $150.1184$, respectively.
Using these values and by using the result in Theorem 1, we then plotted the upper bound on the error vs probability of noise (see Fig.~\ref{fig:err_vs_p_thm1}).
\end{comment}

\begin{figure*}[t!]
    \centering
    \begin{subfigure}[t]{0.5\textwidth}
        \centering
        \includegraphics[scale=0.75]{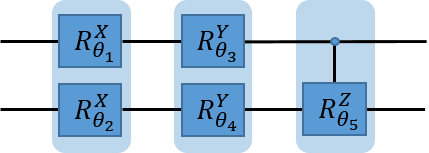}
        \caption{}
        \label{fig:simple_ansatz}
    \end{subfigure}%
    ~ 
    \begin{subfigure}[t]{0.5\textwidth}
        \centering
        \includegraphics[scale =0.75]{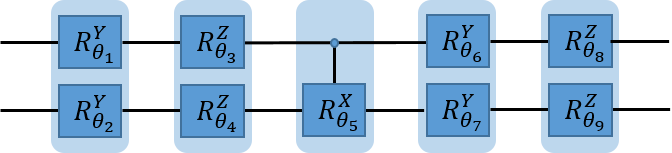}
        \caption{}
        \label{fig:universal_ansatz}
    \end{subfigure}
    \caption{(a) Ansatz with five parameters. (b) Ansatz with nine parameters}
    \label{fig:ansatz}
\end{figure*}

\begin{figure}
    \centering
    \includegraphics[scale=0.6]{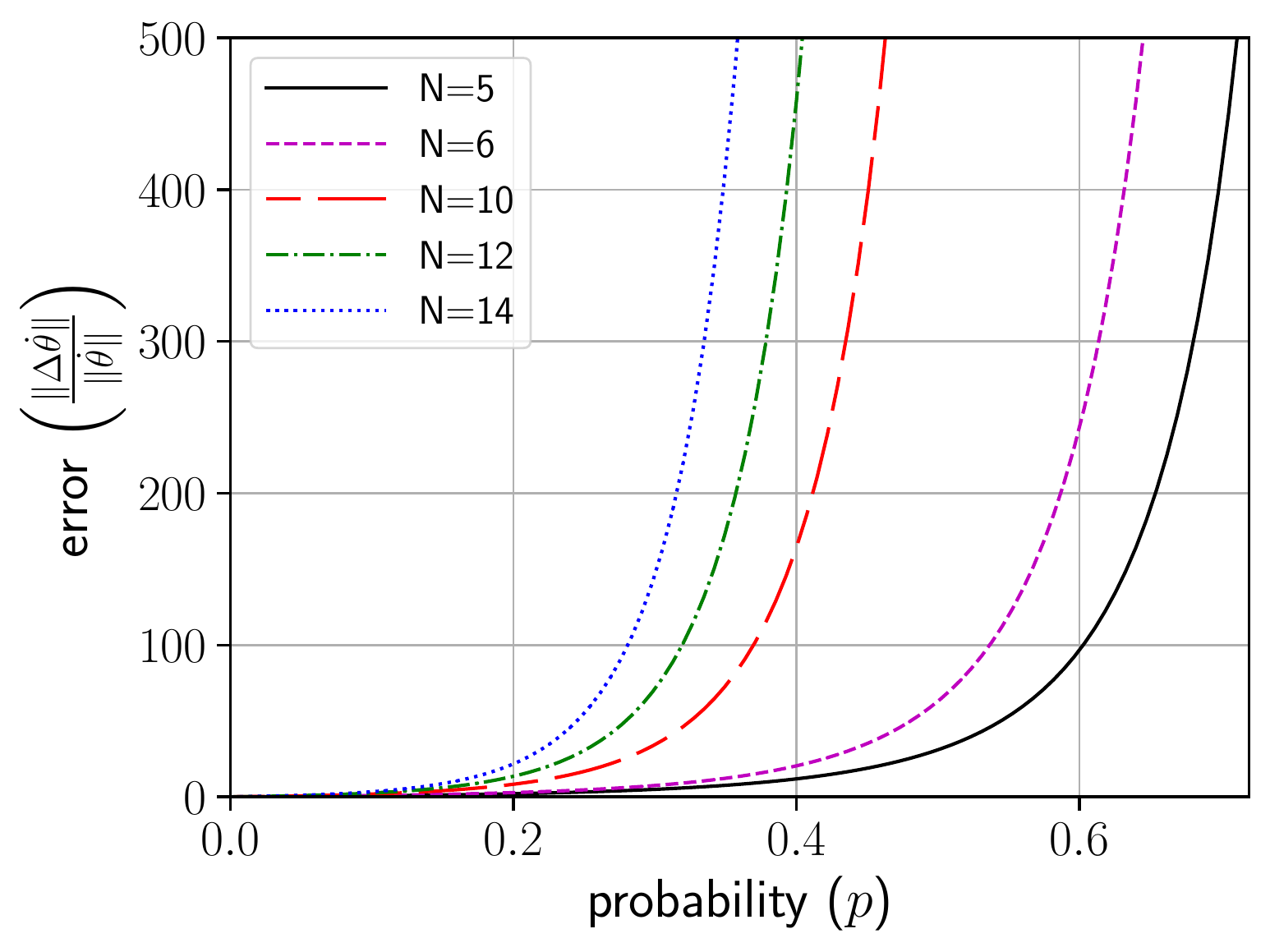}
    \caption{\justifying{Plot of relative error (under partially depolarizing noise) with respect to the probability of depolarizing noise for varying number of parameters, $N$.}}
    \label{fig:err_vs_p_thm2}
\end{figure}

\begin{figure*}[t!]
    \centering
    \begin{subfigure}[t]{0.5\textwidth}
        \centering
        \includegraphics[scale=0.55]{err_vs_p_fiveparams.pdf}
        \caption{}
        \label{fig:err_vs_prob_5params}
    \end{subfigure}%
    ~ 
    \begin{subfigure}[t]{0.5\textwidth}
        \centering
        \includegraphics[scale =0.55]{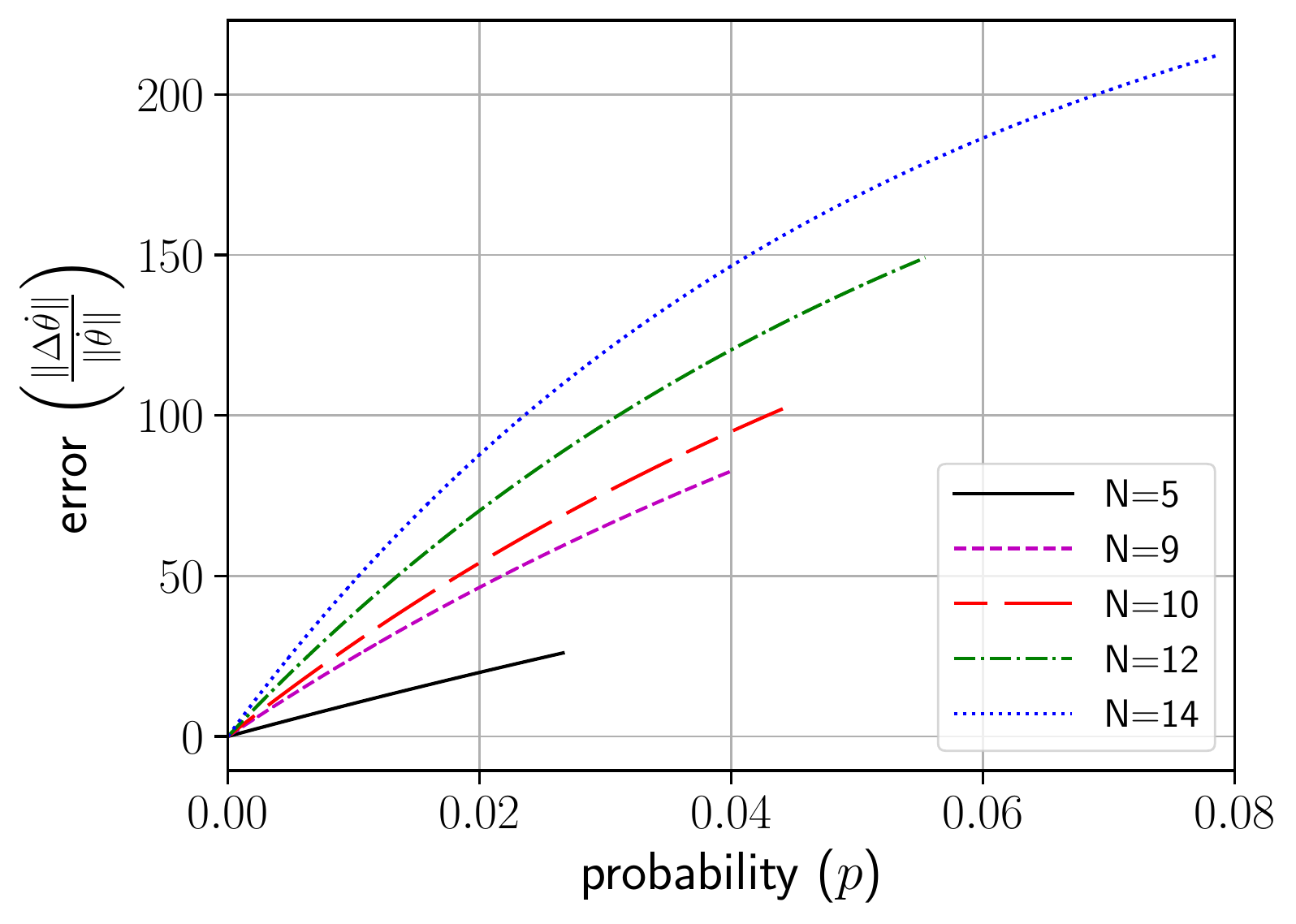}
        \caption{}
        \label{fig:err_vs_prob_9params}
    \end{subfigure}
    \caption{\justifying{Plots of theorem 1 (relative error in the rate of change of parameters with respect to the probability of general probabilistic noise) for two different ansatzes. In each of the above plots, we fixed the condition number of $M$ and varied the number of parameters, $N$. We computed the condition number and the Frobenius norm of $M$ using the five-parameter and the nine-parameter ansatz that we considered for simulating the Hamiltonian of the Hydrogen molecule and by generating random angles. For (a) we fixed the condition number to be $66.7239$ and norm of M to be 0.9977, and for (b), we fixed the condition number to be $20.22$ and the norm of M to be 1.689.
    The curves are plotted up to their valid probabilities as calculated from Eq.~\eqref{eq:constraint_on_prob_tight_ub}.}}
    \label{fig:plots_err_vs_p}
\end{figure*}

Then, by choosing two ansatzes one with five parameters (see Fig.~\ref{fig:simple_ansatz}) and the other with nine parameters (see Fig.~\ref{fig:universal_ansatz}), we again obtained the matrices $M$ and $Y$.
The nine parameter ansatz is the universal ansatz taken from Ref.~\cite{expressibility}.
Using these ansatz and by generating random angles, we obtained $M$, $Y$, $\dot{\theta}$.
For the case when the noise is partially depolarizing, the relative error is independent of the circuit and Hamiltonian, and only depends on $N$ and probability of noise, $p$. The plot between relative error as a function of probability of depolarizing noise for varying $N$ is given in Fig.~\ref{fig:err_vs_p_thm2}.
To obtain the plots of relative error with respect to probability of general hardware noise, we then fixed the condition number and varied the number of parameters. 
The condition number that we used to plot Fig.~\ref{fig:err_vs_prob_5params} was $66.7239$ and the norm of $M$ was 0.9977, both of which were obtained using the ansatz in Fig.~\ref{fig:simple_ansatz} by using the following angles: $\theta_1 = 1.5249,
\,\theta_2 = 2.5142,
\,\theta_3 = 0.4457,
\,\theta_4 = 1.3250,
\,\theta_5 = 2.8769$.
The condition number that we used to plot Fig.~\ref{fig:err_vs_prob_9params} was $20.22$ which was obtained using the ansatz in Fig.~\ref{fig:universal_ansatz} by using the following angles: $\theta_1 = 1.294,\, \theta_2 = 3.265,\,\theta_3= 5.187,\,\theta_4= 3.819,\,\theta_5= 1.5708,\,\theta_6= 4.172,\,\theta_7= 0.618,\,\theta_8= 3.441,\,\theta_9= 2.502$.
The Frobenius norm of $M$ and $Y$ was calculated to be $1.689$ and $0.121$, respectively.
In both these figures, $\delta$ was fixed to be $0.04$; we also fixed the condition number, $\|M\|$ and $\|Y\|$ as obtained above, and varied $N$.
Code/data will be made available upon request.
One can see that for a certain allowed probability of noise, relative error in $\dot{\theta}$ is more for the case with more number of parameters.

\end{document}